\documentclass[a4paper,11pt]{article}
\usepackage{jheppub}
\usepackage[T1]{fontenc}
\usepackage{diagbox}
\usepackage{adjustbox}
\usepackage{multirow}
\usepackage{pifont}
\usepackage{array}
\usepackage{mathtools}
\usepackage{booktabs}
\usepackage[dvipsnames]{xcolor}
\usepackage{tcolorbox}
\usepackage{float}
\usepackage{placeins}
\usepackage{ulem}

\usepackage{verbatim}

\begin{document}

\title{\boldmath  An improved light-cone harmonic oscillator model for the $\phi$-meson longitudinal leading-twist light-cone distribution amplitude}

\author[a]{Dan-Dan Hu}
\emailAdd{hudd@stu.cqu.edu.cn}
\author[a]{, Xing-Gang Wu}
\emailAdd{wuxg@cqu.edu.cn}
\author[a]{, Long Zeng}
\emailAdd{zlong@cqu.edu.cn}
\author[b]{, Hai-Bing Fu}
\emailAdd{fuhb@cqu.edu.cn}
\author[b]{and Tao Zhong}
\emailAdd{zhongtao1219@sina.com}
\affiliation[a]{Department of Physics, Chongqing Key Laboratory for Strongly Coupled Physics, Chongqing University, Chongqing 401331, P.R. China}
\affiliation[b]{Department of Physics, Guizhou Minzu University, Guiyang 550025, P.R.China}

\abstract{In the present paper, we study the properties of $\phi$-meson longitudinal leading-twist light-cone distribution amplitude $\phi_{2;{\phi}}^{\|}(x,\mu)$ by starting from a light-cone harmonic oscillator model for its wavefunction. To fix the input parameters, we derive the first ten $\xi$-moments of $\phi_{2;{\phi}}^{\|}(x,\mu)$ by using the QCD sum rules approach under the background field theory. The shape of $\phi_{2;{\phi}}^{\|}(x,\mu=2~{\rm GeV})$ tends to be a single-peak behavior, which is consistent with the latest Lattice QCD result. As an application, we derive the $D^+_s \to \phi$ transition form factors (TFFs) by using the light-cone sum rules approach. At the large recoil point, we obtain $A_1(0) = 0.512_{-0.020}^{+0.030}$, $A_2(0) = 0.402_{-0.067}^{+0.078}$, $A_0(0) =  0.596_{-0.020}^{+0.025}$ and $V(0) = 0.882_{-0.036}^{+0.040}$. As for the two typical ratios $\gamma_V$ and $\gamma_2$, we obtain $\gamma_V = 1.723_{-0.021}^{+0.023}$ and $\gamma_2 = 0.785_{-0.104}^{+0.100}$. After extrapolating those TFFs to the physically allowable region, we then obtain the transverse, longitudinal and total decay widths for semi-leptonic decay $D^+_s\to\phi\ell^+\nu_{\ell}$. Then the branching fractions are ${\cal B}(D^+_s\to \phi e^+\nu_e) = (2.367_{-0.132}^{+0.256})\times 10^{-3}$ and ${\cal B}(D^+_s\to \phi \mu^+\nu_{\mu}) = (2.349_{-0.132}^{+0.255})\times 10^{-3}$, which show good agreement with the data issued by the BESIII, the CLEO, and the BABAR Collaborations. We finally calculate $D^+_s\to\phi\ell^+ \nu_\ell$ polarization and asymmetry parameters.}

\maketitle

\section{Introduction}

The heavy-to-light semi-leptonic decay is an ideal platform for exploring the properties of the involved heavy quark and the light meson and for testing the Standard Model (SM). It also presents a remarkable opportunity to scrutinize lepton flavor universality (LFU) and to explore the potential manifestations of new physics beyond the SM. For the $D^+_s$ meson semi-leptonic decay, $D^+_s\to \phi\ell^+\nu_{\ell}$, it is important because the $\phi$-meson with $J^{\rm PC}=1^{--}$ is constructed by the $(s\bar{s})$-component within the valance quark model with a narrow resonance that can be well isolated experimentally. Therefore, accurately determining the branching fraction for $D^+_s\to \phi\ell^+\nu_{\ell}$ is crucial for studying the properties of $\phi$ and provides a complementary examination for LFU.

In early 1990s, the Fermilab E687, E791, CLEO and FOCUS Collaborations explored this channel through photoproduction processes~\cite{FermilabE653:1993iut, E687:1994zxt, CLEO:1994msc, E791:1998zxs, FOCUS:2004gfa}. During the past two decades, the BABAR Collaboration presented the branching fraction ${\cal B}(D^+_s\to\phi\ell^+\nu_{\ell}) = (2.61 \pm 0.03 \pm 0.17)\%$ in a relative measurement using a $10~{\rm MeV}$ mass requirement for $\phi\to K^+ K^-$ and taking $D^+_s\to K^+ K^- \pi^+$ as their reference mode~\cite{BaBar:2008gpr}. In 2015, CLEO Collaboration used their measured $D_s^+ \to f_0 e^+ \nu_{e}$ branching fraction and Flatte model-based Monte Carlo simulations to obtain electron channel branching fraction of $D^+_s\to \phi e^+ \nu_e$~\cite{Hietala:2015jqa}. In 2017, the BESIII Collaboration reported the $D^+_s\to \phi \ell^+ \nu_{\ell}$ branching fraction for muon channel~\cite{BESIII:2017ikf}. However, there is a little gap in the branching fractions between the electron and muon channels. Recently in 2023, BESIII updated the absolute branching fraction for $D_s^+\to \phi\mu^+\nu_\mu$~\cite{BESIII:2023opt}. Those data give us good chances for exploring and testing the non-perturbative properties of $D^+_s$ and $\phi$ mesons.

The $D^+_s\to\phi$ transformation are key components for studying the decay channel $D^+_s\to\phi\ell^+\nu_{\ell}$, which can be decomposed into four transformation form factors (TFFs) due to the Lorentz structure of its hadronic matrix element~\cite{Du:2003ja}. These TFFs, which incorporate nonperturbative effects due to the large QCD coupling constant in low $q^2$-region and the bound state effects in large $q^2$-region, have been treated under various approaches. Specifically, the $D^+_s\to\phi$ TFFs have been calculated by using the traditional three-point sum rules (3PSR)~\cite{Du:2003ja, Bediaga:2003hr}, the light-cone sum rules (LCSR) in the framework of heavy quark effective field theory (HQEFT)~\cite{Wu:2006rd, Aliev:2004vf}, the heavy meson chiral Lagrangians (HM$\chi$T)~\cite{Fajfer:2005ug}, the covariant quark model (CQM) \cite{Melikhov:2000yu, Soni:2019huk}, the covariant confining quark model (CCQM) \cite{Ivanov:2019nqd}, the covariant light-front quark model (CLFQM) \cite{Wang:2008ci, Verma:2011yw, Cheng:2017pcq}, the light-front quark model (LFQM)~\cite{Chang:2019mmh}, the lattice QCD (LQCD)~\cite{Gill:2001jp, Donald:2011ff, Donald:2013pea}, the chiral unitary approach ($\chi {\rm UA}$)~\cite{Hussain:1995jq, Sekihara:2015iha}, the relativistic quark model (RQM)~\cite{Faustov:2019mqr}, and the symmetry-preserving regularization of a vector $\times$ vector contact interaction (SCI)~\cite{Xing:2022sor}. Those approaches are valid in different $q^2$-regions. For example, the QCD sum rules (QCD SR) approach, either the LCSR or the 3PSR, is applicable for the low-and-intermediate $q^2$-region, while the LQCD and the HM$\chi$T are applicable for the large $q^2$-region. Predictions under various approaches are complementary to each other~\cite{Huang:2004hw}. And because the LCSR prediction is applicable in a wider region and can be adapted for all $q^2$-region via proper extrapolations, and in this paper, we will adopt the LCSR approach to calculate TFFs. It is noted that the previous LCSR predictions of the $D^+_s\to\phi$ TFFs and their related ratios $r_2$ and $r_V$~\cite{Aliev:2004vf} exhibit significant discrepancies with the experimental measurements. Therefore, it is crucial to recalculate them.

In the LCSR approach, a two-point correlation function is introduced for the heavy-to-light TFFs and expanded near the light cone $x^2\to 0$, whose matrix elements are parameterized as light-cone distribution amplitudes (LCDAs) of increasing twists~\cite{Khodjamirian:2011jp}. The notion of the LCDAs refers to matrix elements of non-local operators sandwiched between the hadron state and the vacuum. Among them, the twist-2 (or equivalently called the leading-twist) LCDAs always gives dominant contributions. It is worth noting that, the TFFs at the large recoil point can be typically affected by ${\cal O}(10\%)$ by the non-asymptotic terms of LCDAs in the LCSR approach, c.f. Refs.~\cite{Ball:1997rj, Ball:2004rg, Ball:2005vx, Wu:2007vi, Bell:2013tfa}. Thus the $s$-quark mass effect need to be taken into consideration for a careful study on the properties of $\phi$-meson. At present, the $\phi$-meson leading-twist LCDA has been calculated by using the QCD SR~\cite{Ball:2007zt}, the Dyson-Schwinger equation (DSE)~\cite{Gao:2014bca}, the Bethe-Salpeter wave functions (BSWF)~\cite{Serna:2022yfp}, and the Algebraic model (AM)~\cite{Almeida-Zamora:2023rwg}, etc. In 2022, the LCDAs of the longitudinally and transversely polarized $\phi$-meson have been calculated by using the lattice QCD based on the large momentum effective theory~\cite{Hua:2020gnw}. At present, various theoretical predictions for the $\phi$-meson's leading-twist LCDAs still exhibit a discrepancy, which also motivate this work.

Generally, the LCDA can be obtained by integrating over the transverse momentum dependence of the light-cone wavefunction (LCWF). In this paper, we will first construct a light-cone harmonic oscillator model (LCHO) model for the $\phi$-meson leading-twist LCWF based on the Brodsky-Huang-Lepage (BHL) description~\cite{Brodsky:1980vj, Brodsky:1981jv, Lepage:1982gd}, which will then be applied to constrain the behavior of $\phi$-meson longitudinal leading-twist LCDA $\phi_{2;\phi}^{\|}$. We will take two ways (resulting to model I and model II, respectively) to fix the input parameters of $\phi_{2;\phi}^{\|}$, both of which need to know the moments of $\phi_{2;\phi}^{\|}$; And we will calculate them by applying the QCD SR within the framework of background field theory (BFTSR). The key idea of the background field theory is to describe the non-perturbative effects with the classical background field satisfying the equation of motion and to describe the quantum fluctuation on this basis within the framework of quantum field theory~\cite{Govaerts:1983ka, Huang:1986wm, Huang:1989gv}. The BFTSR method provides a systematic description of the vacuum condensates from the field theory point of view and a viable way to consider the non-perturbative effects. To take the QCD background field as the starting point for the QCD sum rules, it not only shows a distinct physical picture but also greatly simplifies the calculation due to its capability of adopting different gauge conditions for the quantum fluctuations and backgrounds, respectively. More explicitly, the vacuum expectation values of the background fields well describe the non-perturbative effects, while the quantum fluctuations represent the calculable perturbative effects. As for model I, the least squares method will be firstly employed to fix the $n_{\rm th}$-order moment $\langle\xi^{\|; n}_{2; \phi}\rangle$, which will be directly adopted to determine the input parameters of $\phi_{2;\phi}^{\|}$ by using the definition,
\begin{equation}
\langle \xi_{2;{\phi}}^{\parallel ;n}\rangle {|_\mu } = \int_0^1 {\rm d}x\,{{(2x - 1)}^{n}} \phi_{2;{\phi}}^\| (x,\mu ),
\end{equation}
where $\mu$ represents some initial scale. The model II is followed from the usual idea that the LCDA can be expanded as a Gegenbauer polynomial series, the $n_{\rm th}$-order Gegenbauer moment $a_{2;\phi}^{n}$ will firstly be determined by using the moments of $\phi_{2;\phi}^{\|}$, and then the parameters of $\phi_{2;\phi}^{\|}$ will be fixed by using the following equation,
\begin{align}
a_{2;\phi}^n(\mu)=\frac{\int_{0}^{1} {\rm d}x\,\phi_{2;\phi}^{\|}(x, \mu) C_n^{3/2}(2x-1)}{\int_{0}^{1} {\rm d}x\, 6x(1-x)\left[C_n^{3/2}(2x-1)\right]^2}.
\label{Eq:DA}
\end{align}

The remaining parts of the paper are organized as follows. In Section~\ref{Sec_II}, we describe the calculation technology for deriving the $D^+_s\to\phi$ TFFs and the moments of the $\phi$-meson longitudinal leading-twist LCDA $\phi_{2;\phi}^{\|}$, and for the construction of the LHCO model and the determination of its input parameters. Then we present the detailed numerical analysis and discussion in Section~\ref{Sec_III}. Section~\ref{Sec_IV} is reserved for a brief summary.

\section{Calculation Technology}\label{Sec_II}

\subsection{The $D^+_s \to \phi\ell^+ \nu_\ell$ semi-leptonic decay width}

In the standard model, the matrix element for the semi-leptonic decay $D^+_{s}\to \phi\ell^{+}\nu_{\ell}$ can be written as
\begin{align}
\mathcal{M}(D^+_{s}\to \phi\ell^{+}\nu_{\ell})=\frac{G_F}{\sqrt 2}V_{cs} H^{\mu} L_{\mu},
\end{align}
where $G_F = 1.1663787(6)\times10^{-5} {\rm GeV}^{-2}$ is Fermi constant and $V_{cs}$ is the Cabibbo-Kobayashi-Maskawa (CKM) matrix element for the weak transition $c\to s$. $H^\mu$ and $L_\mu$ represent hadronic transition matrix element and the leptonic current, respectively, which are defined as follows
\begin{equation}
H^\mu =\langle \phi|V^\mu-A^\mu|D_s^+ \rangle, \quad L_\mu=\bar\nu_\ell\gamma_\mu(1-\gamma_5)\ell^+ ,
\end{equation}
with $V^\mu=(\bar q\gamma^\mu c)$ and $A^\mu=(\bar q\gamma^\mu\gamma_5 c)$ stand for the flavor-changing vector and axial-vector currents, respectively. The leptonic part has a simple structure that can be easily calculated using the lepton spinors. The transition between different hadrons is related to bound state effects and hadronization, which has non-perturbation mechanical characteristics. So the hadronic part is much more complicated and requires proper non-perturbative treatment within the QCD theory.

The hadronic matrix element comprises four vectors involved in the transition, namely the four momentum and polarization vectors of the meson, which can be commonly expressed as various parameterized TFFs. More explicitly, the hadronic matrix element for $D^+_s \to \phi$ can be parameterized in terms of five TFFs as follows~\cite{Li:2009tx}:
\begin{eqnarray}
\langle\phi(p,\lambda)|\bar s\gamma_\mu(1- \gamma_5) c|D^+_s(p+q)\rangle
  &=& -i e_\mu^{*(\lambda )}(m_{D^+_s}+ m_\phi)A_1(q^2) \nonumber\\
& & + i(2p+q)_\mu \frac{e^{*(\lambda)} \cdot q}{m_{D^+_s}+ m_\phi} A_2(q^2) \nonumber \\
& &  + iq_\mu (e^{*(\lambda)}\cdot q)\frac{2m_\phi}{q^2}[A_3(q^2)- A_0(q^2)] \nonumber\\
& & - \epsilon^{\mu\nu\alpha\beta}e_\nu^{*(\lambda )}q_\alpha p_\beta \frac{2V({q^2})}{m_{D^+_s} + m_\phi} ,
\label{Eq:TFFp}
\end{eqnarray}
where $m_{D^+_s}$ and $m_\phi$ are masses of $D^+_s$ and $\phi$ mesons, respectively, $p=p_\phi$ is the $\phi$-meson momentum, $q=(p_{D^+_s}-p_\phi)$ is the momentum transfer, and $e^{*(\lambda )}$ stands for the polarization vector of $\phi$-meson with $\lambda = (\bot,\|)$, representing its transverse or longitudinal component, respectively. The TFFs $A_1(q^2)$ and $A_2(q^2)$ are associated with the exchange of a particle with quantum state $J^P=1^+$, while the TFF $V(q^2)$ is associated with $J^P=1^-$~\cite{Richman:1995wm, Becirevic:2020rzi}. The TFF $A_{3}(q^2)$ is not independent, which can be expressed as a linear relation of $A_{1}(q^2)$ and $A_{2}(q^2)$, i.e.
\begin{align}
A_3(q^2)=\frac{(m_{D^+_s}+m_\phi)}{2 m_\phi}A_1(q^2) -\frac{(m_{D^+_s}-m_\phi)}{2 m_\phi}A_2(q^2).
\end{align}

\begin{figure}[htb]
\centering
\includegraphics[width=0.4\textwidth]{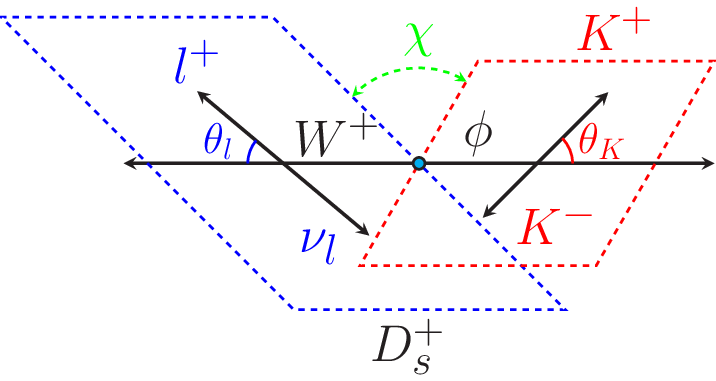}
\caption{The diagram that illustrates the angular configurations utilized in the calculation of the differential rate for the process $D^+_s\to\phi \ell^+ \nu_{\ell}$. The momenta of $\ell^+ \nu_{\ell}$ are depicted in the rest frame of the virtual $W^+$ boson, while those of $K^+K^-$ are represented in the rest frame of $\phi$ meson.}\label{fig:decay}
\end{figure}

The problem of calculating the decay distribution aside, the obtained $D^+_s\to\phi$ TFFs are also linked to the crucial angle variables and provide helpful information to determine their magnitudes. All the final-state particles, excluding the neutrino, can be reconstructed with relatively high efficiency. Four independent kinematic variables characterize the semi-leptonic decay $D^+_s\to\phi\ell^+ \nu_{\ell}$, wherein the $\phi$ meson will subsequently decay into two pseudo-scalar mesons such as $K^+K^-$. The common choices of the four variables are $q^2$ and the three angles $\theta_\ell$, $\theta_K$ and $\chi$ shown in Figure~\ref{fig:decay}, respectively. The polar angle $\theta_\ell$ stands for the angle between the momentum of the charged lepton in the rest frame of the intermediate $W^+$-boson and the direction opposite to the momentum of the final $\phi$-meson in the rest frame of the $D_s^+$-meson. $\theta_K$ is the angle between the momentum of $K^+$ meson and the centre-of-mass momentum of the $K^+K^-$ pair. $\chi$ is the angle between the two planes defined by the $(\ell^+ \nu_{\ell})$ and $(K^+K^-)$ pairs.

To calculate the differential decay rate for $D_s^+\to\phi\ell^+\nu_\ell$, it is convenient to express the hadronic matrix element by the helicity amplitudes $H_{\pm,0,t}$. Those helicity amplitudes are Lorentz-invariant functions which can be formally expressed as the linear combination of the TFFs as follows,
\begin{align}
H_{\pm}(q^2)&=\frac{\lambda^{1/2} (m_{D^+_s}^2, m_\phi^2,q^2)} {m_{D^+_s}+m_{\phi}} \left[\frac{(m_{D^+_s}+m_{\phi})^2}{\lambda^{1/2} (m_{D^+_s}^2,m_\phi^2,q^2)}A_1(q^2)\mp V(q^2)\right],
\nonumber\\
H_0(q^2)&=\frac{1}{2m_{\phi}\sqrt{q^2}}\left[(m_{D^+_s}+m_{\phi})(m^2_{D^+_s}-m^2_{\phi}-q^2) A_1(q^2)-\frac{\lambda (m_{D^+_s}^2, m_\phi^2,q^2)}{m_{D^+_s}+m_{\phi}} A_2(q^2)\right],
\nonumber\\
H_t(q^2)&=\frac{\lambda^{1/2} (m_{D^+_s}^2,m_\phi^2,q^2)}{\sqrt{q^2}}A_0(q^2).
\label{Eq:2}
\end{align}
Main difference between the helicity amplitudes and the usual TFFs lies in the method of decomposition. The $\gamma$-structures of the hadronic matrix elements can be decomposed into Lorentz-invariant structures by using covariant decomposition for the TFFs decomposition method, while the helicity amplitude decomposition method uses the polarization vector of the off-shell $W^+$-boson to obtain a newly combined Lorentz-invariant structure. The use of helicity amplitudes have some advantages in comparison to the usual treatment of TFFs~\cite{Bharucha:2010im, Fu:2014pba, Fu:2014uea, Cheng:2018ouz}: I) Dispersive bounds on the helicity amplitudes parameterization can be achieved via the diagonalizable unitarity relations; II) The polarized decay widths of $D_s^+\to\phi$ can be well studied by using the helicity amplitudes; III) Since the helicity amplitudes have a definite spin-parity quantum number, it is convenient to consider the contribution of the excited state of meson in the transition process; IV) The polarization and asymmetry parameters of $D_s^+\to\phi$ can be accurately determined through the utilization of helicity amplitudes.

By using the helicity amplitudes, the differential decay rate for $D_s^+\to\phi\ell^+\nu_\ell$ with the cascade decay $\phi\to K^+K^-$ can be expressed in terms of the above suggested four kinematic variables as $q^2$, $\theta_\ell$, $\theta_K$ and $\chi$~\cite{Donald:2013pea}, {\it e.g.}
\begin{align}
\frac{{\rm d}\Gamma(D^+_s\to\phi\ell^+ \nu_{\ell},\; \phi\to K^+K^-)}{{\rm d}q^2{\rm d}\cos \theta_K {\rm d}\cos \theta_\ell {\rm d}\chi}&=\frac{3}{8(4\pi)^4}G^2_F|V_{\rm cs}|^2 \frac{|{\bf p_2}| m^2_\ell}{m^2_{D^+_s}}{\cal B}(\phi\to K^+K^-)
\nonumber\\
&\times \big\{\sin^2\theta_K\sin^2\theta_\ell|H_+(q^2)|^2+\sin^2\theta_K\sin^2\theta_\ell|H_-(q^2)|^2
\nonumber\\
&+ 4\cos^2\theta_K\cos^2\theta_\ell|H_0(q^2)|^2+4\cos^2\theta_K|H_t(q^2)|^2
\nonumber\\
&+ \sin^2\theta_K\sin^2\theta_\ell\cos2\chi H_+(q^2)H_-(q^2)
\nonumber\\
&+\sin2\theta_K\sin2\theta_\ell\cos2\chi H_+(q^2)H_0(q^2)
\nonumber\\
&+\sin2\theta_K\sin2\theta_\ell\cos2\chi H_-(q^2)H_0(q^2)
\nonumber\\
&+ 2\sin2\theta_K\sin\theta_\ell\cos\chi H_+(q^2)H_t(q^2)
\nonumber\\
&+ 2\sin2\theta_K\sin\theta_\ell\cos\chi H_-(q^2)H_t(q^2)
\nonumber\\
&+8\cos^2\theta_K\cos\theta_\ell H_0(q^2)H_t(q^2)\big\},
\label{Eq:angle}
\end{align}
where $|{\bf p_2}|=\lambda^{1/2}(m_{D^+_s}^2, m_\phi^2, q^2)/(2m_{D^+_s})$ represents the magnitude of the {\bf 3}-momentum of the daughter meson in the rest frame of $D_s^+$, and $\lambda(x,y,z)=(x+y-z)^2-4xy$ is the K\"{a}llen function. The helicities of the vector meson $\phi$ and the $W^+$ boson must be identical due to the spin-zero nature of the parent meson. The amplitudes for the helicities $0$, $+1$ and $-1$ are proportional to $H_0(q^2)$, $H_+(q^2)$ and $H_-(q^2)$, respectively. Additionally, $H_t(q^2)$ is proportional to $1/\sqrt{q^2}$ and is most significant at low $q^2$-region. The detailed dynamics of the hadronic current is described by the variation of these helicity amplitudes with $q^2$. To obtain the distribution of the polar angle $\theta_\ell$, one can integrate Eq.~\eqref{Eq:angle} over the angle $\theta_K$ and the azimuthal angle $\chi$. Detailed processes for the integration can be found in Ref.~\cite{Ivanov:2019nqd}. And the resultant differential decay width over $q^2$ and $\cos\theta_\ell$ can be expressed ~\cite{Faustov:2019mqr}
\begin{align}
\frac{{\rm d}\Gamma(D_s^+\to\phi\ell^+\nu_\ell)}{{\rm d}q^2{\rm d}\cos \theta_\ell} & = \frac{G^2_F}{(2\pi)^3} |V_{\rm cs}|^2 \frac{\lambda^{1/2} (m_{D^+_s}^2, m_\phi^2, q^2) (q^2 - m^2_\ell)^2}{64 m_{D^+_s}^3 q^2} \left[ (1 + \cos\theta_\ell ) {\cal H}_{\rm U} + 2\sin^2\theta_\ell {\cal H}_{\rm L}  \right. \nonumber\\
& \left. +2\cos \theta_\ell{\cal H}_{\rm P}+\frac{m^2_\ell}{q^2}(\sin^2\theta_\ell {\cal H}_{\rm U}+2\cos^2\theta_\ell{\cal H}_{\rm L}+2{\cal H}_{\rm S}-4\cos \theta_\ell{\cal H}_{\rm SL})\right],
\label{Eq:dgamma}
\end{align}
where the $m_\ell$ with $\ell= (e, \mu)$ is the lepton mass and the helicity structure functions ${\cal H}_{i}$ are defined as
\begin{align}
& {\cal H}_{\rm U} = |H_+|^2 + |H_-|^2, &&{\cal H}_{\rm L} = |H_0|^2,&&{\cal H}_{\rm P}=|H_+|^2-|H_-|^2,
\nonumber\\
& \mathcal{H}_{\rm SL}={\rm Re}(H_0 H_t^\dagger), && \mathcal{H}_{\rm S}=|H_t|^2 .
\end{align}

Based on the helicity amplitudes described above, the forward-backward asymmetry $A^{\ell}_{\rm FB}(q^2)$, lepton-side convexity parameter $C^{\ell}_{\rm F}(q^2)$, and the longitudinal (transverse) polarization of the final charged lepton $P^{\ell}_{\rm L(T)}(q^2)$, as well as the longitudinal (transverse) polarization fraction of the final $\phi$-meson $F^{\ell}_{\rm L(T)}(q^2)$ in the semi-leptonic decay $D_s^+\to \phi\ell^+\nu_\ell$, are expressed as follows~\cite{Faustov:2019mqr}
\begin{eqnarray}
A^{\ell}_{\rm FB}(q^2) &=& \frac{\int^1_0 {\rm d}\cos\theta{\rm d}\Gamma/{\rm d}\cos\theta -\int^0_{-1} {\rm d}\cos\theta {\rm d}\Gamma/{\rm d}\cos\theta } {\int^1_0 {\rm d}\cos\theta {\rm d}\Gamma/{\rm d}\cos\theta +\int^0_{-1} {\rm d}\cos\theta {\rm d}\Gamma/{\rm d}\cos\theta }=\frac{3}{4}\frac{\mathcal{H}_{\rm P}-2\delta\mathcal{H}_{\rm SL}}{\mathcal{H}_{\rm total}}, \label{sys1} \\
C^{\ell}_{\rm F} (q^2) &=& \frac{3}{4}(1-\delta)\frac{\mathcal{H}_{\rm U}-2\mathcal{H}_{\rm L}}{\mathcal{H}_{\rm total}}, \label{sys2} \\
P^{\ell}_{\rm L} (q^2) &=& \frac{(\mathcal{H}_{\rm U}+\mathcal{H}_{\rm L})\left(1-\frac{\delta}{2}\right)-\frac{3\delta}{2}\mathcal{H}_{\rm S}}{\mathcal{H}_{\rm total}}, \label{sys3} \\
P^{\ell}_{\rm T} (q^2) &=& -\frac{3\pi m_\ell}{8\sqrt{q^2}}\frac{\mathcal{H}_{\rm P}+2\mathcal{H}_{\rm SL}}{\mathcal{H}_{\rm total}}, \label{sys4} \\
F^{\ell}_{\rm L} (q^2) &=& \frac{\mathcal{H}_{\rm L}\left(1+\frac{\delta}{2}\right)+\frac{3\delta}{2}\mathcal{H}_{\rm S}}{\mathcal{H}_{\rm total}}.   \label{sys5}
\end{eqnarray}
Here, the total helicity amplitude takes the form $\mathcal{H}_{\rm total}=(\mathcal{H}_{\rm U} + \mathcal{H}_{\rm L}) (1+\delta/2)+3\delta \mathcal{H}_{\rm S} /2$ with $\delta=m^2_\ell/q^2$ and $F^{\ell}_{T}(q^2)=1-F^{\ell}_{L}(q^2)$.

By further integrating over $\cos\theta_\ell$ for differential decay width~\eqref{Eq:dgamma}, we obtain
\begin{align}
\frac{{\rm d}\Gamma(D_s^+\to\phi\ell^+\nu_\ell)}{{\rm d} q^2}=\frac{G_F^2}{(2\pi)^3}|V_{\rm cs}|^2\frac{\lambda^{1/2} (m_{D^+_s}^2,m_\phi^2,q^2)(q^2-m_\ell^2)^2}{24 m_{D^+_s}^3 q^2} \mathcal{H}_{\rm total}
\end{align}
In this paper, the lepton is considered as $\ell=(e, \mu)$ and thus the leptonic mass can be neglected due to chiral suppression. Subsequently, the differential decay width for $D_s^+\to\phi\ell^+\nu_\ell$ can be written as~\cite{Aliev:2004vf, Fu:2018yin, Zhong:2023cyc}
\begin{align}
\frac{{\rm d}\Gamma}{{\rm d} q^2}&=\frac{G_F^2|V_{\rm cs}|^2}{192\pi^3 m_{D^+_s}^3 }\lambda^{1/2} (m_{D^+_s}^2,m_\phi^2,q^2)q^2[|H_+|^2+|H_-|^2+|H_0|^2]
\nonumber\\
&=\frac{{\rm d}\Gamma_{\rm L}}{{\rm d} q^2}+\frac{{\rm d}\Gamma_{\rm T}^+}{{\rm d} q^2}+\frac{{\rm d}\Gamma_{\rm T}^-}{{\rm d} q^2}.
\label{Eq:dg}
\end{align}
The above formula indicates that there are three polarization states for the $\phi$-meson: one longitudinal state and two transverse polarization states (right- and left-handed). The differential decay width for the longitudinally polarized $\phi$-meson has the form
\begin{align}
\frac{{\rm d}\Gamma_{\rm L}}{{\rm d}q^2}& = \frac{{G_F^2|{V_{\rm cs}}{|^2}}}{192\pi^2m_{D^+_s}^3}\lambda^{1/2} (m_{D^+_s}^2,m_\phi^2,q^2) q^2 \bigg|\frac{1}{2m_\phi \sqrt {q^2}}\bigg[(m_{D^+_s}^2 - m_\phi^2 -q^2)(m_{D^+_s}^2 + m_\phi ){A_1}(q^2)
\nonumber\\
& - \frac{{\lambda (m_{D^+_s}^2,m_\phi^2,q^2)}}{m_{D^+_s}^2+m_\phi}{A_2}(q^2)\bigg]{\bigg|^2}, \label{Eq:dgammaL}
\end{align}
and for the transverse differential decay width, we have
\begin{align}
\frac{{\rm d}\Gamma_{\rm T}^{\pm} }{{\rm d}q^2} &= \frac{G_F^2|{V_{\rm cs}}|^2}{192\pi^2m_{D^+_s}^3}\lambda^{1/2} (m_{D^+_s}^2,m_\phi^2,q^2) q^2 \Bigg|(m_{D^+_s} + m_\phi )A_1(q^2)\mp \frac{\lambda ^{1/2}(m_{D^+_s}^2,m_\phi^2,q^2)}{(m_{D^+_s} + m_\phi )}V(q^2)
\Bigg|^2. \label{Eq:dgammaT}
\end{align}
Here the symbols ``$+$'' and ``$-$'' denote the right-handed and left-handed states, respectively. The combined transverse and total decay widths are $\Gamma_{\rm T} = \Gamma_{\rm T}^+  + \Gamma_{\rm T}^-$ and $\Gamma=\Gamma_{\rm total}=\Gamma_{\rm L} + \Gamma_{\rm T}$, respectively. After integrating over $q^2$, we finally obtain the branching fraction of the process
\begin{displaymath}
\mathcal{B}(D_s^+\to\phi\ell^+\nu_\ell)=\tau_{D^+_s} \int^{(m_{D^+_s}-m_\phi)^2}_{m^2_\ell} \frac{{\rm d}\Gamma} {{\rm d}q^2},
\end{displaymath}
where $\tau_{D_s^+}$ is the $D_s^+$-meson lifetime.

\subsection{The $D^+_s \to \phi$ TFFs within the LCSR approach}

In this subsection, we adopt the LCSR approach to calculate the TFFs $A_0(q^2)$, $A_1(q^2)$, $A_2(q^2)$ and $V(q^2)$, which are associated with the current $\bar s(x)\gamma_{\mu} (1-\gamma_5 )c(x)$ and have been defined in Eq.~\eqref{Eq:TFFp}. For the purpose, we adopt the following correlation function to derive the $D_s^+ \to \phi$ TFFs
\begin{align}
\Pi_\mu (p,q) &= -i\int {\rm d}^4x\,e^{iq\cdot x}\langle \phi(p,\lambda) |T\{j_\mu(x),j_{D^+_s}^\dag(0)\}|0\rangle
\nonumber \\
& =  \Pi_1 e_\mu^{*(\lambda )} - \Pi_2(e^{*(\lambda)}\cdot q)(2p+q)_\mu
 - \Pi_3(e^{*(\lambda)}\cdot q){q_\mu } - i{\Pi_V} \epsilon_\mu^{\alpha\beta\gamma} e_\alpha^{*(\lambda )} q_\beta p_\gamma,
\label{Eq:cf}
\end{align}
where $j_\mu(x)=\bar s(x)\gamma_{\mu} (1-\gamma_5 )c(x)$ and $j_{D^+_s}^\dag (x) = \bar c(x)(1 - \gamma_5)s(x)$. To derive the LCSRs for the wanted TFFs, we need to deal with the above correlation function from two aspects. On the one hand, to deal with the correlation function in the spacelike $q^2$-region, we can apply the operator product expansion (OPE) to the right-hand-side of Eq.(\ref{Eq:cf}) near the light cone $x^2\to 0$. The contributions to OPE can be obtained by contracting the quark fields to a full $c$-quark propagator. Processes for dealing with the OPE part is similar to the case of $B\to K^*$ TFFs, which have been given in Ref.\cite{Fu:2014uea}. The interested readers may turn to this reference for detailed calculation technology. Mainly, the matrix elements of the non-local operators between the vector $\phi$-meson and vacuum states can be arranged as the LCDAs with increasing twists, {\it e.g.} leading-twist, twist-3, twist-4 LCDAs, and so on. Contributions from the higher twist LCDAs are generally power suppressed from the leading-twist one, which however may have sizable contributions in certain cases. In our present calculation, we will derive the LCSRs for the TFFs up to twist-4 accuracy. And up to twist-4 accuracy, we have~\cite{Ball:2004rg, Ball:1998ff}
\begin{align}
&\langle 0|\bar s(0) \gamma_\mu s(x)|\phi (p,\lambda )\rangle = m_\phi f_\phi^\| \bigg\{ \frac{e^{*(\lambda )}\! \cdot\! x}{p \cdot x}p_\mu \!\int_0^1 \!{\rm d}u\,e^{iup \cdot x}\bigg[\phi_{2;\phi}^\|(u) + \frac{m_\phi^2 x^2}{16}\phi_{4;\phi}^\|(u)\bigg]
\nonumber\\
&\qquad\qquad\qquad  - \frac12 x_\mu \frac{e^{*(\lambda)} \cdot x}{(p\cdot x)^2} m_\phi^2 \int_0^1 {\rm d}u\, e^{iup\cdot x}\bar g_3(u) + e_{\bot\mu}^{*(\lambda )}\int_0^1 {\rm d}u\, {e^{iup \cdot x}}\phi_{3;\phi }^ \bot (u)\bigg\},
\\
& \langle 0|\bar s(0){\gamma_\mu }{\gamma_5}s(x)|\phi (p,\lambda )\rangle
= - \frac{1}{4} m_\phi f_\phi^\|  {\varepsilon_{\mu \nu \alpha \beta }}e_\nu^{*(\lambda )}{p_\alpha }{x_\beta }\int_0^1 {\rm d}u\,{e^{iup \cdot x}}\psi_{3;\phi }^ \bot (u),
\end{align}
where $\bar g_3 (u)$ is a combined function for short, {\it e.g.} $\bar g_3 (u) =[\psi_{4;\phi}^\|(u) +\phi_{2;\phi}^\|(u) -2\phi_{3;\phi}^\bot(u)]$. $\phi_{2;\phi}^\|(u)$ is the longitudinal leading-twist LCDA, $\phi_{3;\phi}^\bot(u)$ and $\psi_{3;\phi}^\bot(u)$ are twist-3 LCDAs, and $\phi_{4;\phi}^\|(u)$ and $\psi_{4;\phi}^\|(u)$ are twist-4 LCDAs, respectively. Definitions of those LCDAs can be found in Refs.~\cite{Ball:2004rg, Ball:1998ff, Ball:1996tb}.

On the other hand, the correlation function \eqref{Eq:cf} in the timelike $q^2$-region can be treated with the hadronic representation, which is achieved by inserting a complete set of intermediate states with the same quantum numbers in the correlation function. By isolating the pole term of the lowest pseudoscalar $D^+_s$-meson, we obtain the following representation of the correlation function from the hadron side:
\begin{align}
\Pi_\mu(p,q)  &= \frac{{\langle \phi|\bar s \gamma_\mu (1-\gamma_5) c |D^+_s \rangle \langle D^+_s|\bar c \gamma_5 s|0\rangle }}{m_{D^+_s}^2 - (p+q)^2} + \sum\limits_{\rm H}\frac{\langle \phi|\bar s\gamma_\mu (1-\gamma_5) c|{D_s^{\rm H}}\rangle \langle {D_s^{\rm H}}|\bar c(1-\gamma_5)s|0\rangle }{m_{{D_s^{\rm H}}}^{2} - (p+q)^2},
\label{Eq:HR}
\end{align}
where $\langle D^+_s|\bar c \gamma_5 q|0\rangle =i m_{D^+_s}^2 f_{D^+_s}/(m_c+ m_s)$ with $f_{D^+_s}$ being the decay constant of $D^+_s$-meson. To derive the invariant amplitudes $\Pi_{0,1,2,V}$ required in hadronic representation, we adopt the standard sum rules processes. By substituting the parameterization of the matrix elements~\eqref{Eq:TFFp} into~\eqref{Eq:HR}, we obtain
\begin{align}
\Pi_{0,1,2,V}^{\rm had} [q^2,(p+q)^2]&=\frac{m_{D^+_s}^2f_{D^+_s}} {m_c+m_s}
\frac{1}{m^2_{D^+_s} -(p+q)^2}  C_{0,1,2,V}A_{0,1,2,V}(q^2)+\cdots ~,
\end{align}
where the symbol ``$\cdots$'' represents the invariant amplitudes of the high resonance states and the continuum states. The coefficients $C_{0,1,2,V}$ take the values: $C_0 = - 2 m_\phi m^2_{D^+_s}f_{D^+_s}/q^2$, $C_1 = m_{D^+_s}+m_\phi$, $C_2 = - 1/(m_{D^+_s}+m_\phi)$ and $C_V = - 2i/(m_{D^+_s}+m_\phi)$. Then we can write the invariant amplitudes $\Pi^{\rm had}_{0,1,2,V}$ using the general dispersion relation~\cite{Belyaev:1993wp}
\begin{align}
\Pi_{0,1,2,V}^{\rm had}[q^2,(p+q)^2]&=\int_{t_{\rm min}}^\infty\frac{\rho_{0,1,2,V}(q^2,s)}{s-(p+q)^2}{\rm d}s,
\label{Eq:IA}
\end{align}
where possible subtractions are neglected, and the concrete expressions of spectral density $\rho_{0,1,2,V}(q^2,s)$ are provided as follows,
\begin{align}
\rho_{0,1,2,V}(q^2,s)&=\delta(s-m^2_{D^+_s})\frac{m_{D^+_s}^2f_{D^+_s}}{(m_c+m_s)}C_{0,1,2,V}A_{0,1,2,V}(q^2)
 +\rho^{\rm H}_{0,1,2,V}(q^2,s).
\label{Eq:zf}
\end{align}
The second term in Eq.~\eqref{Eq:zf} corresponds to the spectral density of higher resonances and continuum states, which can be approximated by invoking the conventional quark-hadron duality ansatz~\cite{Shifman:1978by}
\begin{align}
\rho^{\rm H}_{0,1,2,V}(q^2,s)=\frac{1}{\pi}{\rm Im}\,\Pi^{\rm OPE}_{0,1,2,V}(q^2,s)\theta(s-s_0)
\label{Eq:da}
\end{align}
where $s_0$ is the threshold parameter and ${\rm Im}\,\Pi^{\rm OPE}_{0,1,2,V}(q^2,s)$ is obtained from the imaginary part of the correlation function~\eqref{Eq:cf} calculated in the OPE. To suppress the contributions from the higher excited states and the continuum states, we apply the usual Borel transformation with respect to the dispersion integration~\eqref{Eq:IA},
\begin{align}
\Pi_{0,1,2,V}[q^2,M^2]&=\int_{t_{\rm min}}^\infty\rho_{0,1,2,V}(q^2,s)e^{-s/M^2}{\rm d}s ,
\label{Eq:BF}
\end{align}
where $t_{\rm min}=(m_c+m_s)^2$. With the help of \eqref{Eq:zf} and \eqref{Eq:da}, we then obtain
\begin{align}
\Pi_{0,1,2,V}[q^2,M^2]&=\frac{m_{D^+_s}^2f_{D^+_s}}{(m_c+m_s)}e^{-m^2_{D^+_s}/M^2} C_{0,1,2,V}A_{0,1,2,V}(q^2)
\nonumber\\
&+\frac{1}{\pi}\int_{s_0}^\infty {\rm Im}\Pi^{\rm OPE}_{0,1,2,V}(q^2,s)e^{-s/M^2} {\rm d}s.
\label{Eq:TFF}
\end{align}
Finally, by equating the correlation functions within different $q^2$-regions yield the desired LCSRs for the $D_s^+\to\phi$ TFFs, e.g.,
\begin{align}
A_1(q^2) &= \frac{2m_c f_\phi^\| m_\phi (m_c + m_s)}{m_{D^+_s}^2 f_{D^+_s} (m_{D^+_s} + m_\phi )} \int_{u_0}^1 \, {\rm d}u \, e^{\frac{m_{D^+_s}^2 - s(u)}{M^2}} \bigg( \frac1u\phi_{3;\phi}^ \bot (u) \,-\, \frac{m_\phi^2}{u^2 M^2}{\bar G}_3(u)\bigg)  \,+\, \frac{2m_c m_\phi^3 f_\phi^\|}{m_{D^+_s}^2{f_{D^+_s}}}
\nonumber\\
&\times \frac{m_c + m_s}{m_{D^+_s} \,+\, m_\phi} e^{\frac{m_{D^+_s}^2 - s(u)}{M^2}}  \, \frac{ {\bar G}_3(u)}{q^2-m_c^2-u^2m_\phi^2}\bigg|_{u\to u_0},
\label{Eq:V1q2}
\end{align}
\begin{align}
A_2(q^2)&= \frac{2m_c f_\phi^\| m_\phi (m_c + m_s)(m_{D^+_s} + m_\phi )}{m_{D^+_s}^2 f_{D^+_s}} \int_{u_0}^1 \frac{{\rm d}u}{M^2} e^{\frac{m_{D^+_s}^2 - s(u)}{M^2}} \bigg( \frac{1}{u^2}{\Phi_{2;\phi}^\|}(u) + \frac{m_c^2m_\phi^2}{4u^4M^4}\Phi_{4;\phi}^\|(u)
\nonumber
\\
&- \frac{\Phi_{3;\phi}^\bot (u)}{u^2}  \,+ \, \frac{m_\phi^2}{u^2 M^2} {\bar G}_3(u)\bigg) \!+\! \frac{2m_c f_\phi^\|m_\phi (m_c + m_s)(m_{D^+_s} + m_\phi)} {m_{D^+_s}^2f_{D^+_s}}\bigg[ \, \frac{\Phi_{2;\phi}^\|(u)} {q^2-m_c^2-u^2m_\phi^2}
\nonumber
\\
&+ \bigg(\frac{m^2_c-q^2}{(q^2-m_c^2-u^2m_\phi^2)^5}m_c^2 m_\phi^2 (m^2_c-2u^2m_\phi^2-q^2)\, \Phi_{4;\phi }^\|(u)\,+\,\frac{u^3 m^2_c m^4_\phi}{(q^2-m_c^2-u^2m_\phi^2)^4}
\nonumber
\\
&\times \frac{d}{du}\Phi_{4;\phi }^\|(u)-\frac{u^2 m^2_c m^2_\phi}{4(q^2 - m_c^2 - u^2 m_\phi^2)^3}~\frac{d^2}{d^2u}~\Phi_{4;\phi }^\|(u)\bigg)-\bigg(\frac{2u^3m^2_\phi}{(q^2-m_c^2-u^2m_\phi^2)^3}{{\bar G}_3}(u)
\nonumber
\\
&+\frac{u^2}{(q^2-m_c^2-u^2m_\phi^2)^2}\frac{d}{du}{{\bar G}_3}(u)\bigg)+\frac{\Phi_{3;\phi }^ \bot (u)}{q^2-m_c^2-u^2m_\phi^2}\bigg]e^{\frac{m_{D^+_s}^2 - s(u)}{M^2}}\bigg|_{u\to u_0},
\label{Eq:V2q2}
\end{align}
\begin{align}
A_0(q^2)&= A_3(q^2)+\frac{q^2 m_c f_\phi^\| m_\phi (m_c + m_s)}{m_{D^+_s}^2 f_{D^+_s}} \int_{u_0}^1 \frac{{\rm d}u}{M^2} e^{\frac{m_{D^+_s}^2 - s(u)}{M^2}} \bigg( \frac{1}{u^2}{\Phi_{2;\phi}^\|}(u) - \frac{m_c^2m_\phi^2}{4u^4M^4}\Phi_{4;\phi}^\|(u)
\nonumber
\\
&- \frac{\Phi_{3;\phi}^\bot (u)}{u^2}  \,- \, \frac{(2-u)m_\phi^2}{u^2 M^2} {\bar G}_3(u)\bigg) \!+\!\frac{q^2 m_c f_\phi^\| m_\phi (m_c + m_s)}{m_{D^+_s}^2 f_{D^+_s}}\bigg[ \, \frac{1} {q^2-m_c^2-u^2m_\phi^2}\Phi_{2;\phi}^\|(u)
\nonumber
\\
&+ \bigg(\frac{m^2_c-q^2}{(q^2-m_c^2-u^2m_\phi^2)^5}m_c^2 m_\phi^2 (m^2_c-2u^2m_\phi^2-q^2) \Phi_{4;\phi }^\|(u)+\frac{u^3 m^2_c m^4_\phi}{(q^2-m_c^2-u^2m_\phi^2)^4}\frac{d}{du}
\nonumber
\\
&\times\Phi_{4;\phi }^\|(u)-\frac{u^2 m^2_c m^2_\phi}{4(q^2 - m_c^2 - u^2 m_\phi^2)^3}\frac{d^2}{d^2u}\Phi_{4;\phi }^\|(u)\bigg)+\bigg(\frac{2(m_c^2+3u^2m_\phi^2-q^2)-2u^3m^2_\phi}{(q^2-m_c^2-u^2m_\phi^2)^3}
\nonumber
\\
&\times{{\bar G}_3}(u)+\frac{u(2-u)}{(q^2-m_c^2-u^2m_\phi^2)^2}\frac{d}{du}{{\bar G}_3}(u)\bigg)+\frac{\Phi_{3;\phi }^ \bot (u)}{q^2-m_c^2-u^2m_\phi^2}\bigg]e^{\frac{m_{D^+_s}^2 - s(u)}{M^2}}\bigg|_{u\to u_0},
\label{Eq:V0q2}
\end{align}
\begin{align}
V(q^2)&= \frac{m_cf_\phi^\|m_\phi (m_{D^+_s} + m_\phi )(m_c + m_s)}{2m_{D^+_s}^2f_{D^+_s}} \int_{u_0}^1 {\rm d}u\, e^{\frac{m_{D^+_s}^2 - s(u)}{M^2}}\, \frac{\psi_{3;\phi }^ \bot (u)}{u^2 M^2}\,-\,\frac{1}{q^2-m_c^2-u^2m_\phi^2}
\nonumber\\
&\times  \psi_{3;\phi }^\bot(u)e^{\frac{m_{D^+_s}^2 - s(u)}{M^2}}\bigg|_{u\to u_0},
\label{Eq:Aq2}
\end{align}
where $u_0=\left(q^2-s_0+m_\phi^2+\sqrt{(q^2-s_0+m_\phi^2)^2-4m^2_\phi(q^2-m_c^2)}\right)/\left(2 m^2_\phi\right)$ and $s(u)=[m_c^2-\bar u(q^2-u m_\phi^2)]/u$ with $\bar u=1-u$. Here, two simplified DAs are defined as
\begin{align}
&\Phi_{2;\phi}^\| (u) = \int_0^u {\rm d}v\, \phi_{2;\phi}^\| (u), \\
&\bar G_3(u) = \int_0^u {{\rm d}v} \int_0^v {\rm d}w\, {{\bar g}_3}(w).
\end{align}
The helicity amplitudes $H_{\pm,0,t}$, the total and differential decay widths can be calculated once the above TFFs have been determined.

\subsection{The $\xi$-moments of $\phi$-meson longitudinal leading-twist LCDA}

The calculation of the form factors for the heavy-to-light transition is primarily influenced by the leading-twist LCDAs of the final state light mesons. Contributions from the higher twist LCDAs are generally power suppressed due to the Borel transformation. To ensure a reasonable distribution amplitude, it is necessary to compute the moments of the corresponding distribution amplitude. The two-particle distribution amplitudes are defined as matrix elements of quark-antiquark gauge invariant non-local operators at light-like separations. The quark composition of $\phi$-meson is $s\bar s$, and its corresponding longitudinal leading-twist LCDA $\phi_{2;{\phi}}^\| (x,\mu)$ is defined as~\cite{Ball:2004rg, Ball:2005vx}
\begin{align}
\langle 0 |{\bar s}(z){\gamma_\mu }s(-z)|{\phi}(q,\lambda )\rangle
&= {m_\phi}{f_\phi^\|}\int_0^1 {\rm d}x\,{e^{i(xz\cdot q - \bar xz\cdot q)}} q_\mu
\frac{{{e^{*(\lambda )}} \cdot z}}{{q \cdot z}}\phi_{2;{\phi}}^\| (x,\mu). \label{Eq:phi2}
\end{align}
The $f^\|_\phi$ is the $\phi$-meson longitudinal decay constant. In the above definitions, $q$ and ${e^{*(\lambda )}}$ are the momentum and polarization vector of $\phi$-meson, respectively. The integration variable $x$ corresponds to the momentum fraction carried by the quark. The polarization vector satisfies the relationship $(e^{ * (\lambda )} \cdot z) /({q \cdot z}) \to 1/{m_\phi}$~\cite{Ball:2004rg}. Performing a series expansion on both sides of Eq.~\eqref{Eq:phi2} at the light-like separation $z_{\mu}$ with $z=0$, we obtain:
\begin{align}
\langle 0|{\bar s}(0)/\!\!\! z  {(iz\cdot\overleftrightarrow D)^{n}}{s}(0)|{\phi}(q,\lambda )\rangle={(z\cdot q)}^{n+1} f_\phi^\| \langle\xi^{n;\|}_{2;\phi}\rangle|_\mu,
\end{align}
where the covariant derivative satisfies the relation $(iz \cdot \overleftrightarrow D)^n=(iz\cdot \overrightarrow D - iz\cdot \overleftarrow D)^n$. In order to determine the sum rules for the moments of the $\phi$-meson longitudinal distribution amplitude, we adopt the following correlation function
\begin{eqnarray}
\Pi_{2;\phi}^{(n,0)}(z,q) &=&i \int {\rm d}^4x\, e^{iq\cdot x}\langle 0 |T\{ J_n(x),J_0^\dagger(0)\}|0\rangle
\nonumber \\
&=&(z\cdot q)^{n+2} I_{2;\phi}^{(n,0)}(q^2), \label{Eq:Correlator1}
\end{eqnarray}
where $J_n(x)= \bar s(x)/\!\!\! z (iz\cdot\overleftrightarrow D)^n s(x)$ and $J_0^\dagger(0)= \bar s(0)/\!\!\! z s(0)$ are interpolating currents. Because of the $G$-parity, only even moments of $\phi_{2;\phi}^\|(x,\mu)$ are non-zero, {\it i.e.} $n=(0,2,4,6,\cdots)$. By combining Feynman rule within the framework of BFTSR, one can apply the OPE for the correlation function~(\ref{Eq:Correlator1}) in deep Euclidean region $q^2 \ll 0$. Then, the correlation function can be expanded into three terms including the quark propagators $S_F^s(0,x)$, $S_F^s(x,0)$ and the vertex operators $(iz\cdot \overleftrightarrow{D})^{n}$, which have been given in our previous work~\cite{Hu:2021zmy}. The $S_F^s(0,x)$ and $S_F^s(x,0)$ represent $s$-quark propagators from 0 to $x$ and $x$ to 0, respectively. When dealing with the Lorentz invariant scalar function $\Pi_{2;\phi}^{(n,0)}(z, q)$, the OPE yields a series of local operators of increasing dimension. The expectation values of these operators in the non-perturbative (physical) vacuum are known as vacuum condensates, whose detailed expression can be found in Refs.~\cite{Zhong:2014jla, Hu:2021zmy}. On the other hand, one can insert a complete set of hadronic states intermediated by $\phi$-meson with the same $J^P$-quantum number into the correlation function and consequently obtain
\begin{align}
\text{Im}\,I_{2;\phi,{\rm Had}}^{(n,0)}(q^2)&=\pi \delta(q^2-m_\phi^2)  (f^\|_\phi)^2 \langle \xi^{\|;n}_{2;\phi} \rangle|_\mu \langle \xi_{2;{\phi}}^{\parallel ;0}\rangle
+\text{Im}\,I_{2;\phi}^{\rm pert}(s)\theta(s-s_\phi), \label{rm}
\end{align}
where $s_\phi$ is the continuum threshold. Finally, a detailed expression for the moments of the distribution amplitudes can be obtained by equating the OPE results with the hadronic representation using the dispersion relation and then performing a Borel transformation. And the sum rules for the moment of $\phi$-meson leading-twist LCDA is given by:
\begin{eqnarray}
&&\frac{(f^\|_\phi)^2 \langle\xi^{\|;n}_{2;\phi}\rangle|_\mu \langle \xi^{\|;0}_{2;\phi}\rangle|_\mu}{M^2 e^{m_\phi^2/M^2}}  = \frac{1}{\pi M^2} \int_{4m_s^2}^{s_\phi} {\rm d}s\, e^{-s/M^2}\text{Im} I_{2;\phi}^{\rm pert}(s) + \frac{2m_s\langle\bar ss\rangle}{M^4} + \frac{\langle \alpha_s G^2\rangle}{12\pi M^4}\frac{1+n\theta(n-2)}{n+1}
\nonumber
\\
&&\qquad\qquad - \frac{m_s\langle g_s\bar s\sigma TGs\rangle}{9M^6}\,(8n+1) \,+\, \frac{\langle g_s \bar ss\rangle }{81M^6}4(2n + 1)\,-\,\frac{\langle g_s^3fG^3\rangle} {48\pi^2M^6}n\theta(n-2) \,+\,\frac{\sum\langle g_s^2\bar qq\rangle^2}{486\pi^2M^6}
\nonumber
\\
&&\qquad\qquad \times \bigg\{-2\,(51n+25) \,\bigg(-\ln\frac{M^2}{\mu^2}\bigg)\,+\, 3(17n+35)  \,+\, \theta(n-2)\bigg[2n\bigg(-\ln\frac{M^2}{\mu^2}\bigg)-25
\nonumber
\\
&&\qquad\qquad \times (2n+1)\tilde \psi (n)+ \frac1n(49n^2+100n+56)\bigg]\bigg\}+ m_s^2 \bigg\{\!-\!\frac{\langle\alpha_s G^2\rangle}{6\pi M^6}\bigg[\theta(n-2)(n\tilde\psi (n)-2)
\nonumber
\\
&&\qquad\qquad -n\!-\!2+2n\bigg(\!-\ln\frac{M^2}{\mu^2}\bigg)\bigg]+\frac{\langle g_s^3fG^3\rangle}{288\pi^2M^8}\bigg\{\!-\!10 \delta^{n0} + \!\theta (n \!- \!2) \bigg[ 4n(2n-1)\!\left(\! -\ln\frac{M^2}{\mu^2}\right)
\nonumber
\\
&&\qquad\qquad- 4n\tilde \psi (n) + 8({n^2} - n + 1)\! \bigg] \!+ \theta(n-4)\Big[2n(8n-1)\tilde\psi(n) - (19{n^2} + 19n + 6)\Big]+ 8n
\nonumber
\\
&&\qquad\qquad \times (3n - 1)\left(\! -\ln\frac{M^2}{\mu^2} \right) -\! (21n^2+53n-6)\!\bigg\} \!-\! \frac{\sum \langle g_s^2q\bar q\rangle^2 }{972\pi^2M^8}\bigg\{ 6\delta^{n0}\!\left[ {16\!\left( -\ln\frac{M^2}{\mu^2}\! \right)\! -\! 3} \right]
\nonumber
\\
&&\qquad \qquad+\theta (n - 2) \bigg[8(n^2+ 12n -12)~\left( -\ln\frac{M^2}{\mu^2} \right)-2(29n + 22) \tilde \psi (n)+ 4\bigg(5{n^2} - 2n-33
\nonumber
\\
&&\qquad\qquad  +\frac{46}{n}\bigg)\bigg] + \theta (n - 4)\,\bigg[2\left(56{n^2} - 25n + 24\right) \tilde \psi(n)- \left(139 n^2 + 91n + 54\right)\bigg] +8\,\bigg(27n^2
\nonumber
\\
&&\qquad\qquad -15\,n -11\bigg)\, \bigg(-\ln\frac{M^2}{\mu^2} \bigg) - 3(63n^2+159n-50) \bigg\}\,+\frac{4(n-1)}3 \frac{m_s\langle\bar ss\rangle }{M^6} \,+ \frac{8n-3}9
\nonumber
\\
&&\qquad\qquad \times \frac{m_s\langle g_s\bar s\sigma TGs\rangle}{M^8}- \frac{4(2n+1)}{81}\frac{\langle g_s \bar ss\rangle^2}{M^8}\bigg\},
\label{Eq:xi2}
\end{eqnarray}
where the imaginary part of the perturbative contribution is
\begin{align}
{\rm Im}\,I_{2;\phi}^{\rm pert}(s)&=\frac{3 v^{n+1}}{8\pi(n+1)(n+3)}\bigg\{[1+(-1)^n](n+1)
\frac{1-v^2}2+[1+(-1)^n]\bigg\},
\end{align}
where $v^2=1-4m_s^2/s$ and $\tilde \psi(n)=\psi(\frac{n+1}{2})-\psi(\frac{n}{2})+ \ln 4$. The $0_{\rm th}$-order derivative of the digamma function is given by $\psi(n+1)=\sum_{k=1}^n 1/k-\gamma_E$ with the Euler constant $\gamma_E=0.577216$. Eq.~\eqref{Eq:xi2} indicates that the zeroth order $\xi$-moment $\langle \xi^{\|;0}_{2;\phi}\rangle|_\mu$ cannot be normalized in the entire Borel parameter region due to the truncation of contributions from vacuum condensates with dimensions greater than six. The following equation is employed to enhance the accuracy and rationality of the calculation of the result $\langle\xi^{\|; n}_{2; \phi}\rangle|_\mu$~\cite{Zhong:2014jla}.
\begin{align}
\langle\xi^{\|;n}_{2;\phi}\rangle|_\mu=\frac{\langle\xi^{\|;n}_{2;\phi}\rangle|_\mu\langle\xi^{\|;0}_{2;\phi}\rangle|_\mu
\big|_{{\rm From~Eq}.~(\ref{Eq:xi2})}}{\sqrt{(\langle\xi^{\|;0}_{2;\phi} \rangle|_\mu)^2}}.
\end{align}
Here the squared zeroth order moment $\langle\xi^{\|;n}_{2;\phi}\rangle|_\mu$ in the denominator is obtained by taking $n\to 0$ in Eq.~\eqref{Eq:xi2}, and its detailed expression is given by
\begin{align}
&(\langle \xi^{\|;0}_{2;\phi}\rangle|_\mu)^2= \frac{e^{m_\phi^2/M^2}}{(f^\|_\phi)^2} \int_{4m_s^2}^{s_\phi} {\rm d}s\, e^{-s/M^2}\frac{v}{8\pi^2}(3-v^2) + \frac{2m_s\langle\bar ss\rangle}{M^4} \, + \frac{\langle \alpha_s G^2\rangle}{12\pi M^4} - \frac{m_s\langle g_s\bar s\sigma TGs\rangle}{9M^6}
\nonumber
\\
&\qquad\qquad \,+ \frac{4 \langle g_s \bar ss\rangle }{81M^6} \,+\frac{\sum\langle g_s^2\bar qq\rangle^2}{486\pi^2M^6} \bigg[-50\bigg(-\ln\frac{M^2}{\mu^2}\bigg)+105 \bigg] \,+ m_s^2 \bigg\{\frac{\langle\alpha_s G^2\rangle}{3\pi M^6}-\frac{\langle g_s^3fG^3\rangle}{72\pi^2M^8}
\nonumber
\\
&\qquad\qquad - \frac{\sum \langle g_s^2q\bar q\rangle^2 }{972\pi^2M^8}\bigg [8 \bigg(-\ln\frac{M^2}{\mu^2} \bigg)+132\bigg]-\frac{4}{3} \frac{m_s\langle\bar ss\rangle }{M^6}-\frac{m_s\langle g_s\bar s\sigma TGs\rangle}{3 M^8}- \frac{4}{81}\frac{\langle g_s \bar ss\rangle^2}{M^8}\bigg\},
\label{xi0}
\end{align}
The moments $\langle\xi^{\|;n}_{2;\phi}\rangle|_\mu$ and the Gegenbauer moments $a_{2;\phi}^n(\mu)$ at the same scale $\mu$ are related via the following equations~\cite{Fu:2018vap}:
\begin{align}
\langle\xi^{\|;2}_{2;\phi}\rangle|_\mu &=\frac{1}{5}+\frac{12}{35}a_{2;\phi}^2(\mu)  \label{momentsrel1}\\
\langle\xi^{\|;4}_{2;\phi}\rangle|_\mu &=\frac{3}{35}+\frac{8}{35}a_{2;\phi}^2(\mu)+\frac{8}{77}a_{2;\phi}^4(\mu) \label{momentsrel2} \\
\langle\xi^{\|;6}_{2;\phi}\rangle|_\mu &=\frac{1}{21}+\frac{12}{77}a_{2;\phi}^2(\mu)+\frac{120}{1001}a_{2;\phi}^4(\mu)+\frac{64}{2145}a_{2;\phi}^6(\mu)  \label{momentsrel3} \\
& \cdots \nonumber
\end{align}

\subsection{The $\phi$-meson longitudinal leading-twist LCDA from the LCHO model}

The LCHO model is based on the BHL prescription and the Melosh-Wigner transformation, where the Melosh-Wigner transformation relates the light-cone spin state to the ordinary instant-form spin state wave functions and is one of the most important ingredients of the light-cone formalism~\cite{Melosh:1974cu}. In this context, we can construct a light-cone wave function for the quark-antiquark Fock state in the light-cone quark model by using the Melosh-Wigner rotation~\cite{Yu:2007hp}. And the complete light-front wave function is accomplished by appraising the spin and momentum space wave functions, $\chi^{\Lambda}_{\lambda_1, \lambda_2}(x, {\bf k}_\bot)$ and $\psi_{2;\phi }^{\rm R}(x, {\bf k}_\bot)$, e.g.
\begin{align}
\Psi_{2;\phi}^\| (x,{\bf k}_\bot)= {\chi^{\Lambda}_{\lambda_1, \lambda_2}}(x,{\bf k}_\bot)\psi_{2;\phi}^{\rm R}(x,{\bf k}_\bot),
\end{align}
where $\chi^{\Lambda}_{\lambda_1, \lambda_2}(x, {\bf k}_\bot)$ depends on the $\phi$-meson spin projection. The Fock expansion of the two particle Fock-state for $\phi$-meson includes two different types of spin configurations: longitudinal (L) and transverse (T), each with distinct $\lambda_1$ and $\lambda_2$. Here $\lambda_1$ and $\lambda_2$ represent the helicities of the quark and antiquark, respectively. Within the light front holographic model, the Lorentz invariant spin structure of vector mesons is expressed as~\cite{Kaur:2020emh}
\begin{align}
\chi^{\rm L(T)}_{\lambda_1, \lambda_2}(x,{\bf k}_\bot)=\frac{{\bar u}_{\lambda_1}(k^+,{\bf k}_\bot)}{\sqrt{x}}(\epsilon_{\Lambda}\cdot\gamma)\frac{\nu_{\lambda_2}(k^{\prime +},{\bf k}^{\prime}_\bot)}{\sqrt{1-x}},
\end{align}
where $\epsilon_{\Lambda}$ is the polarization vector, $k$ and $k'$ stand for the 4-momenta of the quark and the antiquark, respectively. The longitudinal and transverse polarizations of vector mesons are given by
\begin{align}
\epsilon_{\rm L} = \bigg(\frac{P^+} {M_V} , -\frac{M_V} {P^+}, 0, 0 \bigg)~~~~~~ \epsilon^{\mp}_{\rm T} = \mp \frac{1}{\sqrt{2}}(0,0,1,\pm i),
\end{align}
where $P$ is the meson momentum and $M_V$ is the mass of the vector meson. The spin wave functions of pseudo-scalar meson and vector meson are derived from the light-front holographic model~\cite{Kaur:2020emh} and the light-cone quark model~\cite{Lu:2007sg, Dhiman:2017urn}. The spin part of the wave function is determined by the Melosh-Wigner method, which establishes a connection between spin states transforming from the instanton form to the light-front form. For the case of $\Lambda={\rm L}$, the spin part of the wave function for the $\phi$-meson reads~\cite{Kaur:2020emh}
\begin{align}
\chi^{\rm L}_{+,+}(x,{\bf k}_\bot) &= + \frac{(1-2x){\cal M}(k_x-ik_y)}{({\cal M} + 2m_s)\sqrt {2({\bf k}_\bot^2 + m_s})},  \\
\chi^{\rm L}_{+,-}(x,{\bf k}_\bot) &= + \frac{m_s({\cal M}+2m_s)+2{\bf k}_\bot^2}{({\cal M} + 2m_s)\sqrt{2({\bf k}_\bot^2 + m_s}) }, \\
\chi^{\rm L}_{-,+}(x,{\bf k}_\bot) &= + \frac{m_s({\cal M}+2m_s)+2{\bf k}_\bot^2}{({\cal M} + 2m_s)\sqrt{2({\bf k}_\bot^2 + m_s}) }, \\
\chi^{\rm L}_{-,-}(x,{\bf k}_\bot) &= - \frac{(1-2x){\cal M}(k_x+ik_y)}{({\cal M} + 2m_s)\sqrt {2({\bf k}_\bot^2 + m_s})},
\end{align}
where $m_s$ is the mass of the constitute $s$-quark in $\phi$-meson and abbreviation $\bar x = (1-x)$ is used. This treatment agrees with the cases of $\rho$-meson due to the same vector meson spin projection~\cite{Qian:2008px}. In the present paper, our main concern is the longitudinal distribution amplitude of the vector meson. Thus we only need to consider the following spin part of the wave function that gives sizable contribution to the present case:
\begin{align}
\chi_{2;\phi}(x,{\bf k}_\bot) = \frac{{m_s({\cal M} + 2m_s) + 2{\bf k}_\bot^2}}{({\cal M} + 2m_s)\sqrt{2({\bf k}_\bot^2 + m_s)}},
\end{align}
where ${\cal M} = \sqrt{({\bf k}_\bot^2 + m_s^2)/(x\bar x)}$ is the invariant mass of the composite system. According to the BHL prescription, the momentum space wave function of $\phi$-meson is~\cite{Guo:1991eb}:
\begin{align}
\psi_{2;\phi}^{\rm R}(x,{{\bf k}_\bot }) = A_{2;\phi}^\| {\varphi^\|_{2;\phi}}(x)\exp \bigg[ - \frac{1}{{8\beta_{2;\phi}^2}}\bigg(\frac{{{\bf k}_\bot^2 + m_s^2}}{x\bar x}\bigg)\bigg],
\end{align}
where $A_{2;\phi}^\|$ and $\beta_{2;\phi}$ are the normalization constant and harmonic parameter, respectively. In the context, the function $\varphi_{2;\phi}^\| (x)$ dominates longitudinal distribution, which can be expressed in the following two formulations~\cite{Zhong:2021epq}
\begin{align}
&\varphi_{2;\phi}^{\|({\rm I})}(x) = 1 + b_{2;\phi}^2 C_2^{3/2}(2x-1)+ b_{2;\phi}^4 C_4^{3/2}(2x-1), \\
&\varphi_{2;\phi}^{\|({\rm II})}(x) = (x\bar x)^{\alpha_{2;\phi}}[1 + B_{2;\phi}^2 C_2^{3/2}(2x-1)].  \label{phimodelII}
\end{align}
The first one, denoted as $\varphi_{2;\phi}^{\|({\rm I})}$, adopts the usual Gegenbauer expansion. Its model parameters $b_{2;\phi}^2$ and $b_{2;\phi}^4$ can be fixed by using the $n_{\rm th}$-order Gegenbauer moments $a_{2;\phi}^n$ that have been defined in Eq.~\eqref{Eq:DA}. To derive a more accurate representation of $\phi_{2;\phi}^\|$, it is necessary to increase the precision of the higher-order Gegenbauer polynomial. This treatment is discussed in detail in our previous work~\cite{Zhong:2014fma}. However, due to large coefficients that exist between $\langle\xi^{n;\|}_{2;\phi}\rangle$ and $a_{2;\phi}^n$, the reliability of calculating $a_{2;\phi}^n$ using the QCD SR will decrease with the increment of the $n_{\rm th}$-order. Thus, we will also adopt another model, namely $\varphi_{2;\phi}^{\|({\rm II})}$, to enhance the form of $\phi_{2;\phi}^\|$, whose longitudinal part explicitly contains a factor $(x\bar x)^{\alpha_{2;\phi}}$ that is close to asymptotic form $\phi_{2;\phi}^\|(x, \mu\to\infty)=6x\bar x$~\cite{Lepage:1980fj}.

By combining the spin and space wave functions, one can obtain the comprehensive wave function of the $\phi$-meson, {\it i.e.}
\begin{align}
\Psi_{2;\phi}^\|(x,{\bf k}_\bot) &= A_{2;\phi}^\| \varphi^\|_{2;\phi}(x) \frac{m_s({\cal M} + 2m_s) + 2{\bf k}_\bot^2}{({\cal M} + 2m_s)\sqrt{2({\bf k}_\bot^2 + m_s})} \exp \left[ - \frac{1}{{8\beta_{2;\phi}^2}}\bigg(\frac{{\bf k}_\bot^2 + m_s^2}{x\bar x}\bigg)\right] .
\end{align}
By utilizing the relationship between $\phi$-meson leading-twist LCDA and the wave function, one then derive the expression for $\phi$-meson longitudinal leading-twist LCDA
\begin{align}
\phi_{2;\phi}^\| (x,\mu)&=\frac{2\sqrt6}{f_\phi^\|} \int_{|{\bf k}_\bot^2| \leq \mu^2}\frac{{\rm d}^2{\bf k}_\bot}{16\pi^3}A_{2;\phi}^\| \varphi^\|_{2;\phi}(x)
 \frac{m_s({\cal M} + 2m_s) + 2{\bf k}_\bot^2}{({\cal M} + 2m_s)\sqrt{2({\bf k}_\bot^2 + m_s})}
\nonumber\\
&\times \exp \left[ - \frac{1}{{8\beta_{2;\phi}^2}}\bigg(\frac{{\bf k}_\bot^2 + m_s^2}{x\bar x}\bigg)\right] .
\end{align}

The next step is to determine the four model-dependent parameters. For $A_{2;\phi}^\|$ and $\beta_{2;\phi}$, they can be constrained by the following two conditions:
\begin{itemize}
\item The normalization of the wave function,
    \begin{align}
       \int_0^1 {\rm d}x \int \frac{{\rm d}^2{\bf k}_\bot}{16\pi^3} \Psi_{2;\phi}^\|(x,{\bf k}_\bot) = \frac{f_\phi^\|}{2\sqrt 6 }.
       \label{Eq:nc}
	\end{align}
\item The probability of finding the lowest Fock state $|s\bar s\rangle$ in the $\phi$-meson expansion,
   	\begin{align}
    	P_\phi &= \int_0^1 {\rm d}x \int \frac{{\rm d}^2{\bf k}_\bot}{16\pi^3}|\Psi_{2;\phi}^\|(x,{\bf k}_\bot)|^2.
   	\label{Eq:qq}
   	\end{align}
\end{itemize}
In this paper, the probability value is chosen as $P_\phi\approx 0.6$, mainly relying on the prediction for $K$-meson $P_K\approx 0.52$ as proposed by Guo and Huang~\cite{Guo:1991eb}. Subsequently, the remaining two parameters, $\alpha_{2;\phi}$ and $B_{2;\phi}^2$ for $\varphi_{2;\phi}^{\|({\rm II})}$, can be fitted by equating its $\xi$-moments $\langle \xi_{2;{\phi}}^{\parallel ;n}\rangle|_\mu$ to the derived values give by the QCD SR (\ref{Eq:xi2})~\cite{Hu:2021lkl}. And to make their values more accurately, we shall adopt the $\xi$-moments up to $10_{\rm th}$-order level.

\section{Numerical results and discussions}\label{Sec_III}

\subsection{Input parameters}

To do the numerical analysis on the properties of $\phi$-meson longitudinal leading-twist LCDA and $D_s^+\to\phi$ TFFs, the values of parameters are taken as follows. The $\phi$-meson mass and decay constant are $m_\phi=1.019~{\rm GeV}$ and $f^\|_\phi=0.231\pm0.004~{\rm GeV}$~\cite{Ball:2004rg}, respectively. The charm-quark current mass $m_c({\bar m}_c)=1.27\pm0.02~{\rm GeV}$, the $s$-quark current mass $m_s(2{\rm GeV})=0.093^{+0.011}_{-0.005}~{\rm GeV}$ and the $D_s^+$-meson mass and decay constant are $m_{D_s^+}=1.968~{\rm GeV}$~\cite{Zyla:2022zbs} and $f_{D^+_s}=0.256\pm0.004~{\rm GeV}$~\cite{Duplancic:2015zna}. When calculating the moments of the distributed amplitude, we also need to know the values of non-perturbative vacuum condensates up to $6$-dimension, which can be read from Ref.~\cite{Zhong:2021epq}. The renormalization scale is set as $\mu_k=(m^2_{D^+_s}-{\bar m}_c^2)^{1/2}\approx 1.5~{\rm GeV}$.

Furthermore, we also need to know the values of the non-perturbative vacuum condensates up to dimension-six~\cite{Zhong:2021epq}:
\begin{align}
&\langle s\bar s\rangle = (-1.789_{-0.084}^{+0.168})\times  10^{-2}~{\rm GeV}^3,
\nonumber\\
&\langle g_s\bar qq\rangle^2 = (2.082_{-0.697}^{+0.734})\times  10^{-3} ~{\rm GeV}^6,
\nonumber\\
&\langle g_s\bar s\sigma TGs\rangle =(-1.431_{-0.076}^{+0.139})\times  10^{-2}~{\rm GeV}^5,
\nonumber\\
&\langle g_s^2\bar qq\rangle^2 = (7.420_{-2.483}^{+2.614})\times 10^{-3}~{\rm GeV}^6,
\nonumber\\
&\langle \alpha_s G^2 \rangle = 0.038\pm0.011~{\rm GeV}^4,
\nonumber\\
&\langle g_s^3fG^3\rangle \approx 0.045 ~{\rm GeV}^6,
\nonumber\\
&\sum \langle g_s^2\bar qq\rangle^2 = (1.891_{-0.633}^{+0.665})\times 10^{-2}~{\rm GeV}^6.
\end{align}
The ratio $\kappa = \langle s\bar s\rangle/\langle q\bar q\rangle= 0.74\pm0.03$ is given in Ref.~\cite{Narison:2014wqa}. To ensure the consistency of the calculation, all vacuum condensates and quark mass should be evolved from some hadronic scale $\mu_0$ that is of order ${\cal O}(1\; {\rm GeV})$ to the required renormalization scale by using the renormalization group equation~(RGE), {\it e.g.}~\cite{Yang:1993bp, Hwang:1994vp, Lu:2006fr, Zhang:2021wnv}.
\begin{align}
&{\bar m}_s(\mu) ={\bar m}_s(\mu_0)\bigg[\frac{\alpha_s(\mu_0)}{\alpha_s(\mu)}\bigg]^{4/\beta_0},
\nonumber\\
&\langle q\bar q\rangle(\mu) = \langle q\bar q\rangle(\mu_0) \bigg[\frac{\alpha_s(\mu_0)}{\alpha_s(\mu)}\bigg]^{-4/\beta_0},
\nonumber\\
&\langle g_s\bar qq\rangle^2(\mu) = \langle g_s\bar qq\rangle^2\bigg[\frac{\alpha_s(\mu_0)}{\alpha_s(\mu)}\bigg]^{-2/(3\beta_0)},
\nonumber\\
&\langle g_s\bar q\sigma TGq\rangle(\mu) = \langle g_s\bar q\sigma TGq\rangle(\mu_0)\bigg[\frac{\alpha_s(\mu_0)}{\alpha_s(\mu)}\bigg]^{-2/(3\beta_0)},
\nonumber\\
&\langle g_s^2\bar qq\rangle^2(\mu)=\langle g_s^2\bar qq\rangle^2(\mu_0),
\nonumber\\
&\langle \alpha_s G^2 \rangle(\mu) = \langle \alpha_s G^2 \rangle(\mu_0),
\nonumber\\
&\langle g_s^3fG^3\rangle(\mu) =\langle g_s^3fG^3\rangle(\mu_0).
\end{align}
Besides, the evolution equation of the Gegenbauer moments is
\begin{align}
a_{2;\phi}^n(\mu)= a_{2;\phi}^n(\mu_0)\bigg[\frac{\alpha_s(\mu)}{\alpha_s(\mu_0)}\bigg]^{\gamma_n/{2\beta_0}},
\label{Eq:gm}
\end{align}
The evolution equation can be expressed as $c_i(\mu_k) = L^{\gamma_n/2\beta_0} c_i(\mu_0)$, where $L = \alpha_s(\mu_k)/\alpha_s(\mu_0)$, $\beta_0 = 11 - 2/3 n_f$ and $n_f$ represents the number of flavors involved. The one-loop anomalous dimension $\gamma_n$ can be expressed in two ways: longitudinal $\gamma_n^\|$ and transverse $\gamma_n^\bot$. In the context, we primarily utilize the longitudinal one-loop anomalous dimension to satisfy the following equation~\cite{Ball:2004rg},
\begin{eqnarray}
\gamma_n^\|  = 8{C_F}\bigg[\psi (n + 2) + {\gamma_E} - \frac{3}{4} - \frac{1}{2(n + 1)(n + 2)}\bigg]
\end{eqnarray}
with $C_F=4/3$. Using Eq.\eqref{Eq:gm} and the linear relationship between the $\xi$-moments $\langle\xi^{\|;n}_{2;\phi}\rangle|_{\mu}$ and Gegenbauer moments $a_{2;\phi}^n(\mu)$, one can derive the $\xi$-moments at any scale $\mu$.

\subsection{Determination of the $\phi$-meson leading-twist LCDA}
\begin{figure}[t]
\centering
\includegraphics[width=0.60\textwidth]{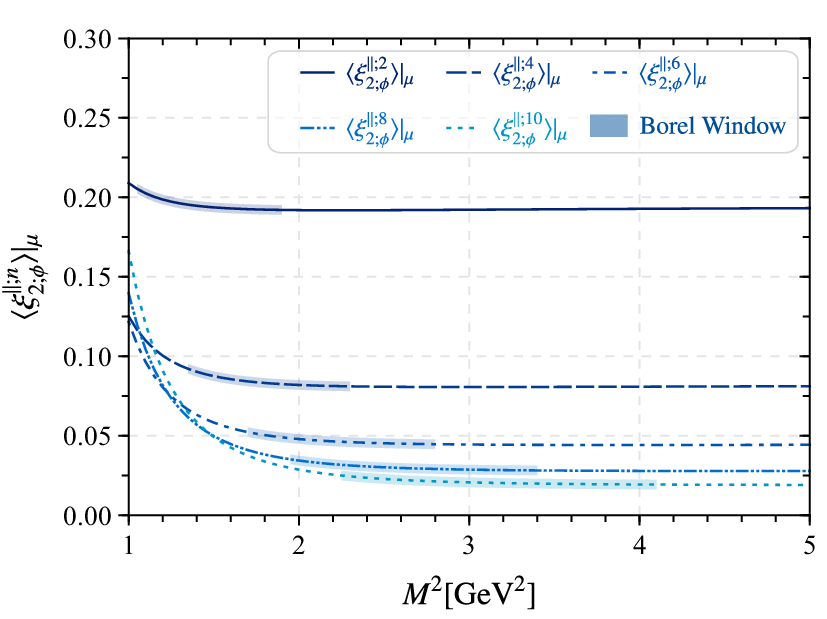}
\caption{Moments of the leading-twist LCDA $\langle\xi^{\|;n}_{2;\phi}\rangle|_{\mu=\sqrt{M^2}{\rm GeV}}$ up to $n=(2,4,6,8,10)$ order level versus the Borel parameter $M^2$. The shaded bands stand for the corresponding Borel windows.}\label{fig:xi}
\end{figure}

According to previous experiences, two crucial parameters need to be established when determining the moments of LCDA within the framework of BFTSR. Within an appropriate Borel window, the $0_{\rm th}$-order Gegenbauer moment is normalized to determine the continuum threshold, namely $s_\phi=2.1{\rm GeV^2}$. When determining the Borel parameters, it is necessary for the contributions from the continuum states and the six-dimensional condensates to be sufficiently small. Numerically, we have found that reasonable Borel windows can be achieved by constraining the contributions from the continuous states for the first five moments are $20\%$, $25\%$, $30\%$, $35\%$ and $40\%$, respectively; and the contributions from the six-dimensional condensation for all moments are not more than $5\%$.

As a conservative prediction, we stipulate that the variations of $\langle\xi^{\|;n}_{2;\phi}\rangle|_{\mu}$ within the Borel window should be less than $10\%$. Figure~\ref{fig:xi} shows how the first five non-zero moments of $\phi$-meson longitudinal leading-twist LCDA at $\mu=\sqrt{M^2}~{\rm GeV}$ change with the Borel parameter $M^2$. The shaded bands represent the corresponding Borel windows, each falling within the region of $[1.0, 5.0]~{\rm GeV}^2$. At the scale $\mu_0=1~{\rm GeV}$, we have
\begin{align}
\langle\xi^{\|;2}_{2;\phi}\rangle|_{\rm 1 GeV} &=0.199\pm0.010 , \label{ximoments1} \\
\langle\xi^{\|;2}_{4;\phi}\rangle|_{\rm 1 GeV} &=0.086\pm0.006 , \label{ximoments2} \\
\langle\xi^{\|;2}_{6;\phi}\rangle|_{\rm 1 GeV} &=0.049\pm0.004 , \label{ximoments3} \\
\langle\xi^{\|;2}_{8;\phi}\rangle|_{\rm 1 GeV} &=0.032\pm0.003 , \label{ximoments4} \\
\langle\xi^{\|;2}_{10;\phi}\rangle|_{\rm 1 GeV} &=0.022\pm0.003. \label{ximoments5}
\end{align}
Using the relationships (\ref{momentsrel1}, \ref{momentsrel2}, \ref{momentsrel3}, $\cdots$) among $\langle\xi^{\|;n}_{2;\phi}\rangle$ and $a_{2;\phi}^n$, the values of $a_{2;\phi}^n$ at the scale $\mu_0=1$ GeV can be obtained accordingly. By applying the evolution equation, the values of $\langle\xi^{\|;n}_{2;\phi}\rangle$ and $a_{2;\phi}^n$ can be obtained at any scale. Using those values together with the constraints listed in previous section, we are ready to fix the $\phi$-meson longitudinal leading-twist LCDA at any scales.
\begin{figure}[htb]
\begin{center}
\includegraphics[width=0.60\textwidth]{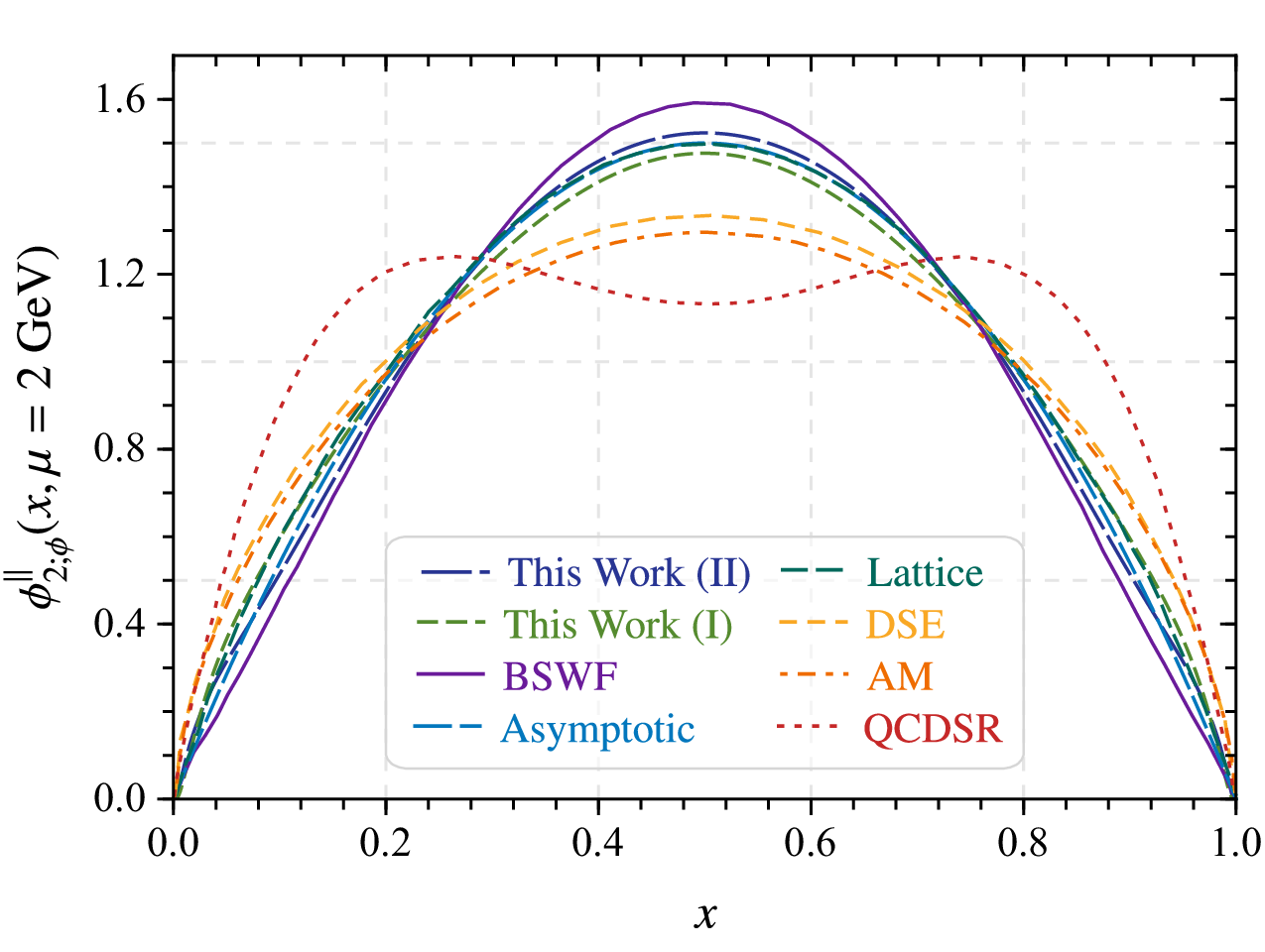}
\end{center}
\caption{The $\phi$-meson longitudinal leading-twist LCDA $\phi^\|_{2;\phi}(x,\mu)$ at the scale $\mu=2~{\rm GeV}$. As a comparison, the results derived from the QCD SR~\cite{Ball:2007zt}, the DSE~\cite{Gao:2014bca}, the BSWF~\cite{Serna:2022yfp}, the AM~\cite{Almeida-Zamora:2023rwg}, the Lattice QCD~\cite{Hua:2020gnw} approaches have been presented. The usual asymptotic behavior~\cite{Lepage:1980fj} for $\mu\to\infty$ is also presented.}
\label{fig:phi}
\end{figure}

We present the $\phi$-meson leading-twist LCDA in Figure~\ref{fig:phi}. As a comparison, we have displayed the results derived from the QCD SR~\cite{Ball:2007zt}, the DSE~\cite{Gao:2014bca}, the BSWF~\cite{Serna:2022yfp}, the AM~\cite{Almeida-Zamora:2023rwg}, the Lattice QCD~\cite{Hua:2020gnw} approaches in Figure~\ref{fig:phi}. And for easy comparison, all the curves are for $\mu=2$ GeV. When $\mu=2~{\rm GeV}$, the corresponding input parameters for our present model I and II are $A^{\|({\rm I})}_{2;\phi}=11.508~{\rm GeV^{-1}}$, $\beta^{({\rm I})}_{2;\phi}=1.213~{\rm GeV}$, $b_{2;\phi}^2=0.061$, $b_{2;\phi}^4=0.020$ and $A^{\|({\rm II})}_{2;\phi}=2.515~{\rm GeV^{-1}}$, $\alpha_{2;\phi}=-0.940$, $B_{2;\phi}^2=-0.149$, $\beta^{({\rm II})}_{2;\phi}=1.207~{\rm GeV}$, respectively. The usual asymptotic behavior has also been presented in Figure~\ref{fig:phi}. Previous QCD SR calculation prefers a double humped behavior~\cite{Ball:2007zt}. Figure~\ref{fig:phi} shows that our present model I and model II are close in shape, both of which prefer the asymptotic form. Numerically, we will find that the TFFs and hence the related observables are almost the same for those two models. Such single-peak behavior is also consistent with the predictions of the Lattice QCD, the DSE, the BSWF, the AM approaches. Thus in the following discussions, we will adopt the model II to do our discussion.
\begin{figure}[htb]
\begin{center}
\includegraphics[width=0.60\textwidth]{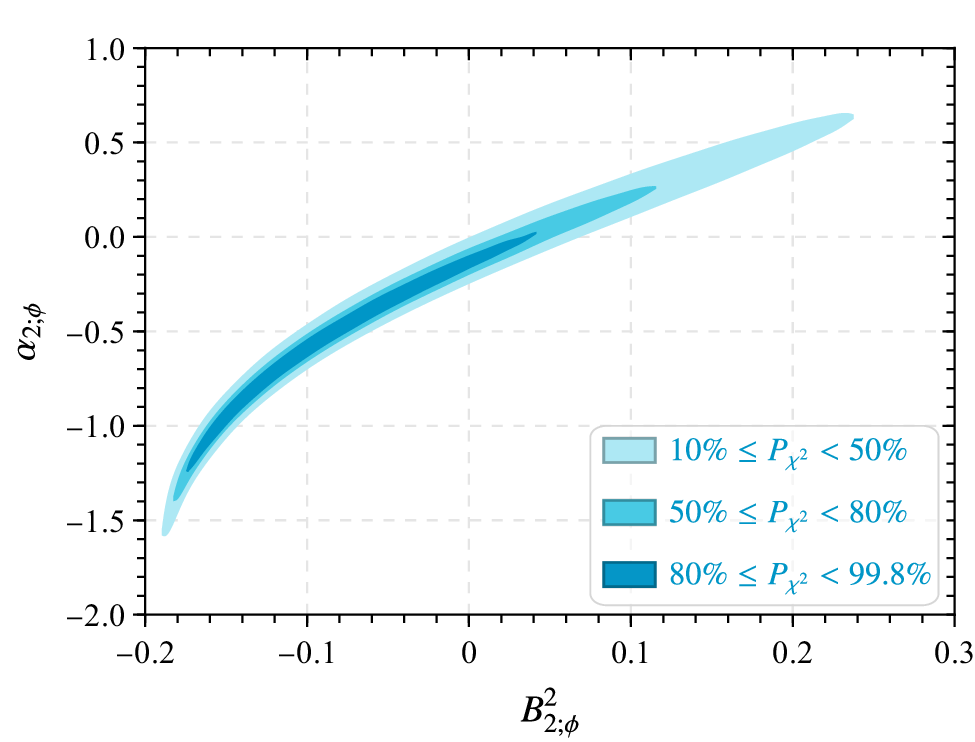}
\end{center}
\caption{The allowable regions for the combined parameters $\alpha_{2;\phi}$ and $B_{2;\phi}^2$ versus the goodness-of-fit $P_{\chi^2}$ for the case of $s$-quark constituent quark mass $m_s=370~{\rm MeV}$. }
\label{fig:fit}
\end{figure}

As mentioned above, the typical scale for the $D_s^+\to\phi$ TFFs is $\mu_k\approx 1.5~{\rm GeV}$. For later convenience, we give the fitting parameters for $\phi^{({\rm II})}_{2;\phi}$ at the scale $\mu_k$, {\it e.g.} $A^{\|({\rm II})}_{2;\phi}=5.808~{\rm GeV^{-1}}$, $\alpha_{2;\phi}=-0.650$, $B_{2;\phi}^2=-0.110$ and $\beta^{({\rm II})}_{2;\phi}=0.989~{\rm GeV}$, respectively. Different $s$-quark constituent quark mass will lead to different input parameters, here we have implicitly set $m_s=370~{\rm MeV}$, which has been derived within the invariant meson mass scheme by Refs.~\cite{Jaus:1991cy, Choi:1997iq}. Under the spin-averaged meson mass scheme, Refs.~\cite{Dziembowski:1986dr, Choi:1996mq} suggest that $m_s=450~{\rm MeV}$. If using this value, we observe that the input parameters will be changed accordingly, which however only slightly changes the shape of $\phi^{({\rm II})}_{2;\phi}$, and hence slight changes to the $D_s^+\to\phi$ TFFs. So we will adopt $m_s=370~{\rm MeV}$ to do our calculations, which also results in a higher goodness-of-fit up to $99.6\%$. Practically, the probability magnitude $P_{\chi^2} (P_{\chi^2} \in [0, 1])$ is employed to judge the goodness-of-fit~\cite{Zyla:2022zbs}, when its value is closer to 1, a better fit is assumed to be achieved. To illustrate the relationship more clearly for the combined magnitudes of $\alpha_{2;\phi}$ and $B_{2;\phi}^2$ with the goodness-of-fit $P_{\chi^2}$, we present the allowable regions for the combined parameters $\alpha_{2;\phi}$ and $B_{2;\phi}^2$ versus the goodness-of-fit $P_{\chi^2}$ in Figure~\ref{fig:fit}. In Figure~\ref{fig:fit}, the darker shaded band represents a better goodness-of-fit. It can be observed that as the goodness-of-fit goes up, the allowable range for the combined parameters goes down.

\subsection{The $D_s^+\to \phi$  TFFs}
\begin{table}[htb]
\renewcommand\arraystretch{1.15}
\footnotesize
\begin{center}
\caption{Comparison of various theoretical predictions for the $D_s^+\to \phi$ TFFs $V(0)$, $A_0(0)$, $A_1(0)$ and $A_2(0)$ at the large recoil point $q^2=0$.} \label{Tab:TFF}
\begin{tabular}{lllll}
\hline
~~~~~~~~~~~~~~~~~~~~~~~~~~~~~~& $V(0)$~~~~~~~~~~~~~~~~~~& $A_0(0)$~~~~~~~~~~~~~~~~~&$A_1(0)$~~~~~~~~~~~~~~~~~& $A_2(0)$   \\   \hline
This~work   & $ 0.882_{-0.036}^{+0.040}$  &$ 0.596_{-0.020}^{+0.025}$ & $ 0.512_{-0.020}^{+0.030}$ & $ 0.402_{-0.067}^{+0.078}$    \\
LQCD'01~\cite{Gill:2001jp}    & $0.85(14)$   & $0.63(2)$ & $0.63(2)$  & $0.62(78)$       \\
LQCD'11~\cite{Donald:2011ff}   & $0.903(67)$   & $0.686(17)$  & $0.594(22)$  & $0.401(80)$       \\
LQCD'13~\cite{Donald:2013pea}  & $1.059(124)$   & $0.706(37)$  & $0.615(24)$  & $0.457(78)$      \\
HQEFT~\cite{Wu:2006rd}    & $0.778_{-0.062}^{+0.057}$ &$\!\!\!\!\!-0.757_{-0.039}^{+0.029}$  & $0.569_{-0.049}^{+0.046}$  & $0.304_{-0.017}^{+0.021}$     \\
${\rm HM\chi T}$~\cite{Fajfer:2005ug}    & $1.10$   & $1.02$   & $0.61$    & $0.32$         \\
CQM~\cite{Melikhov:2000yu}    & $1.10$   & $0.73$   & $0.64$   & $0.47$        \\
3PSR~\cite{Du:2003ja}   & $1.21(33)$  & $0.53(12)$  & $0.55(15)$     & $0.59(11)$         \\
CLFQM'08~\cite{Wang:2008ci}  & $0.91$  & $0.62$      & $0.61$     & $0.58$         \\
CLFQM'11~\cite{Verma:2011yw} & $0.98$  & $0.72$      & $0.69$     & $0.57$     \\
LFQM~\cite{Chang:2019mmh}    & $1.24$  & $0.71$      & $0.77$     & $0.66$     \\
LCSR~\cite{Aliev:2004vf}    & $0.70(10)$  & $0.53(9)$  & $0.54(9)$  & $0.57(9)$        \\
CCQM~\cite{Ivanov:2019nqd}  & $0.91$    & $0.68$   & $0.68$     & $0.67$         \\
RQM~\cite{Faustov:2019mqr}  & $0.999$   & $0.713$  & $0.643$    & $0.492$          \\
SCI~\cite{Xing:2022sor}     & $1.00$   & $0.66$   & $0.61$    & $0.44$           \\
\hline
\end{tabular}
\end{center}
\end{table}

The continuum threshold $s_0$ and the Borel parameter $M^2$ are two important input parameters for the LCSR analysis of the $D_s^+\to \phi$ TFFs. Usually, the continuum threshold $s_0$ is taken as the value that is close to the squared mass of the first excited state of $D^+_s$. As a conservative prediction, we set $s_0^{A_0} = s_0^{A_1} = s_0^{A_2} = s_0^V  = 6.70(25)~{\rm GeV}^2$ to do our discussion, whose central value is set by $D^+_{s0}(2590)$~\cite{Zyla:2022zbs}. The Borel parameter has to be chosen within the certain ``window'' to ensure the best stability of the physical results. The requirement for selecting stable ``window'' is as follows: the Borel parameter can not be excessively large, as higher resonance and continuum state contributions cannot be effectively suppressed; simultaneously, they should not be too small, as the truncated OPE would fail~\cite{Ball:2005vx}. Therefore we chose a slightly flatter section of the Borel window and the determined Borel parameters are ${M}^2_V =2.7(1)~{\rm GeV}^2$, ${M}^2_{A_0} =2.5(2)~{\rm GeV}^2$, ${M}^2_{A_1} = 2.2(2)~{\rm GeV}^2$ and ${M}^2_{A_2} = 2.3(1)~{\rm GeV}^2$, respectively. Using those parameters and the LCSRs \eqref{Eq:V1q2}-\eqref{Eq:Aq2}, we calculate the $D_s^+\to \phi$ TFFs and give their values at the large recoil point, {\it i.e.} $q^2 = 0~{\rm GeV}^2$ in Table~\ref{Tab:TFF}, in which the results given by several groups have also been presented.
\begin{table}[t]
\renewcommand\arraystretch{1.15}
\footnotesize
\begin{center}
\caption{Comparison of various theoretical predictions for the ratio $\gamma_V$ and $\gamma_2$ of $D_s^+\to \phi$ with the available experimental data.} \label{Tab:ratio}
\begin{tabular}{lll}
\hline
Experiments  ~~~~~~~~~~~~~~~~~~~~~~~& $\gamma_V$  ~~~~~~~~~~~~~~~~~~~~~~~~~~~~~~~~~~~~~~~~~& $\gamma_2$   \\
\hline
PDG~\cite{Zyla:2022zbs}       & $1.80\pm0.08$  & $0.84\pm0.11$      \\
BESIII'23~\cite{BESIII:2023opt}  & $1.58\pm0.17\pm0.02$  & $0.71\pm0.14\pm0.02$      \\
BABAR~\cite{BaBar:2008gpr}       & $1.807\pm0.046\pm0.065$  & $0.816\pm0.036\pm0.030$ \\
FOCUS~\cite{FOCUS:2004gfa}      & $1.549\pm0.250\pm0.148$  & $0.713\pm0.202\pm0.284$   \\ \hline
Theories    & $\gamma_V$& $\gamma_2$   \\ \hline
This~work   & $ 1.723_{-0.021}^{+0.023}$     & $0.785_{-0.104}^{+0.100}$      \\
LQCD'01~\cite{Gill:2001jp}       & $1.37(7)$  & $0.98(8)$   \\
LQCD'11~\cite{Donald:2011ff}     & $1.52(12)$    & $0.68(12)$   \\
LQCD'13~\cite{Donald:2013pea}    & $1.72(21)$    & $0.74(12)$   \\
HQEFT~\cite{Wu:2006rd}    & $1.37_{-0.021}^{+0.024}$    & $0.53_{-0.006}^{-0.010}$     \\
${\rm HM\chi T}$~\cite{Fajfer:2005ug}        & $1.80$     & $0.52$     \\
CQM~\cite{Melikhov:2000yu}         & $1.72$        & $0.73$      \\
3PSR~\cite{Du:2003ja}            & $2.20(85)$  & $1.07(43)$      \\
CLFQM'08~\cite{Wang:2008ci}     & $1.49$    & $0.95$      \\
CLFQM'11~\cite{Verma:2011yw}    & $1.42$   & $0.83$  \\
LFQM~\cite{Chang:2019mmh}      & $1.61$   & $0.86$  \\
LCSR~\cite{Aliev:2004vf}      & $1.19(23)$  & $1.06(24)$      \\
CCQM~\cite{Ivanov:2019nqd}     & $1.34(27)$  & $0.99(20)$      \\
RQM~\cite{Faustov:2019mqr}    & $1.56$  & $0.77$       \\
SCI~\cite{Xing:2022sor}       & $1.64$  & $0.72$       \\
\hline
\end{tabular}
\end{center}
\end{table}

Two ratios of the $D_s\to\phi$ TFFs are usually studied, which are defined as:
\begin{align}
{\gamma_V} = \frac{V(0)}{{A_1}(0)},~~~~~~{\gamma_2} = \frac{{A_2}(0)}{{A_1}(0)}.
\end{align}
Our predicted values of $\gamma_V$ and $\gamma_2$ are given in Table~\ref{Tab:ratio}. Additionally, we have included results from other theoretical groups for comparison, including LQCD~\cite{Gill:2001jp, Donald:2011ff, Donald:2013pea}, LCSR~\cite{Aliev:2004vf}, HQEFT~\cite{Wu:2006rd}, ${\rm HM\chi T}$~\cite{Fajfer:2005ug}, CQM~\cite{Melikhov:2000yu}, 3PSR~\cite{Du:2003ja}, CLFQM~\cite{Wang:2008ci, Verma:2011yw}, LFQM~\cite{Chang:2019mmh}, CCQM~\cite{Ivanov:2019nqd}, RQM~\cite{Faustov:2019mqr} and SCI~\cite{Xing:2022sor}. We have also included results from different experimental collaboration groups: PDG~\cite{Zyla:2022zbs}, BESIII'23~\cite{BESIII:2023opt}, BABAR~\cite{BaBar:2008gpr} and FOCUS~\cite{FOCUS:2004gfa}. The predicted TFFs $V(0)$, $A_0(0)$, $A_1(0)$ and $A_2(0)$ show variations across different theoretical groups. Our predictions indicate a high level of consistency between the TFFs $V(0)$ and $A_2(0)$ and the results from LQCD'11, whereas $A_1(0)$ and $A_0(0)$ exhibit smaller values compared to those obtained in LQCD'11. Table~\ref{Tab:ratio} also shows that our numerical results of the two ratios fall within the error range of LQCD'13, BESIII'23 and PDG.

\begin{table}[t]
\renewcommand\arraystretch{1.15}
\centering
\footnotesize
\caption{The masses of low-lying $D_s$ resonances, the coefficients $\alpha_{1,i}$ and $\alpha_{2,i}$, and $\Delta_i$ for the TFFs $V(q^2)$, $A_0(q^2)$, $A_1(q^2)$ and $A_2(q^2)$, respectively. All the input parameters have been set to be their central values. } \label{Tab:SSE}
\begin{tabular}{lllll}
\hline
~~~~~~~~~   & $V(q^2)$~~~~~~~~   & $A_0(q^2)$~~~~~~~~  & $A_1(q^2)$~~~~~~~~   & $A_2(q^2)$   \\   \hline
$m_{R^*,i}$     & 2.1121      & 2.4595        & 2.4595      & 2.4595       \\
$\alpha_{1,i}$    & $-11.852$   & $-5.031$      & $-4.034$   & $-2.562$     \\
$\alpha_{2,i}$    & $354.663$   & $125.030$     & $87.102$     & $11.467$      \\
$\Delta_{i} $      & $0.076\%$   & $0.045\%$     & $0.033\%$     & $0.009\%$    \\
\hline
\end{tabular}
\end{table}

The LCSR approach is applicable in low and intermediate $q^2$-regions, {\it e.g.} $q^2\in [0, 0.54~{\rm GeV}^2]$, and we need to extrapolate the TFFs to all allowable physical region so as to derive the wanted values for the physical observables such as the decay widths and so on. In the present paper, we adopt the approach of simplified series expansion (SSE)~\cite{Bharucha:2015bzk, Bharucha:2010im} to do the extrapolation. One of the advantages of this parametrization is the simplicity to translate the near-threshold behavior of the form factors into a useful constraint on the expansion coefficients. So the TFFs take the following form:
\begin{eqnarray}
F_i(q^2) = \frac{1}{1-q^2 / m^2_{R^*, i}}\sum\limits_{k=1}^2 \alpha_{k,i} z^k(t, t_0).
\end{eqnarray}
Here $F_i(q^2)$ with $i=(1,\cdots,4)$ represents the four TFFs $V(q^2)$ and $A_{0,1,2}(q^2)$, respectively. The function $z(t=q^2, t_0)$, which incorporates the parameters $t_\pm$, $t_0$ and $t$, is defined as
\begin{align}
z(t, t_0)=\frac{\sqrt{t_{+} - t}-\sqrt{t_{+}-t_0}}{\sqrt{t_{+} - t}+\sqrt{t_{+}-t_0}},
\end{align}
where
\begin{align}
t_{\pm}=(m_{D^+_s}\pm m_{\phi})^2, ~~~t_0=t_{+}(1-\sqrt{1-t_{-}/t_{+}}).
\end{align}
In this approach, the simple pole $(1-q^2 / m^2_{R^*, i})$ is used to account for the low-lying resonances, and $m_{R^*, i}$ are $D_s^+$-meson resonances. The masses of the low-lying $D_s^+$ resonances are mainly determined by their $J^P$-states, whose values can be found in Refs.~\cite{Zyla:2022zbs, Momeni:2020zrb}. The free parameters $\alpha_{1,i}$ and $\alpha_{2,i}$ are fixed to make the $\Delta_{i}$ as small as possible, such as $\Delta_{i} <1\%$, where $\Delta_{i}$ is used to measure the quality of extrapolation and is defined as
\begin{align}
\Delta_{i}  = \frac{\sum\nolimits_t|F_i(t)-F_i^{\rm fit}(t)|}{ \sum\nolimits_t |F_i(t)|}\times  100,
\end{align}
where $t \in [0, 1/40, \cdots ,40/40]\times  0.54~{\rm GeV}^2$. We present the masses of the low-lying $D_s^+$ resonances, the fitting parameters $\alpha_i$ for each TFFs and the quality-of-fit $\Delta$ in Table~\ref{Tab:SSE}. It shows that under those choices of parameters, all the $\Delta_{i}$ values of $D_s^+\to\phi$ TFFs are no more than 0.076$\%$, indicating a good agreement of the extrapolated curves with the LCSRs within the same $q^2$-region of $q^2\in [0, 0.54~{\rm GeV}^2]$.
\begin{figure}[htb]
\begin{center}
\includegraphics[width=0.45\textwidth]{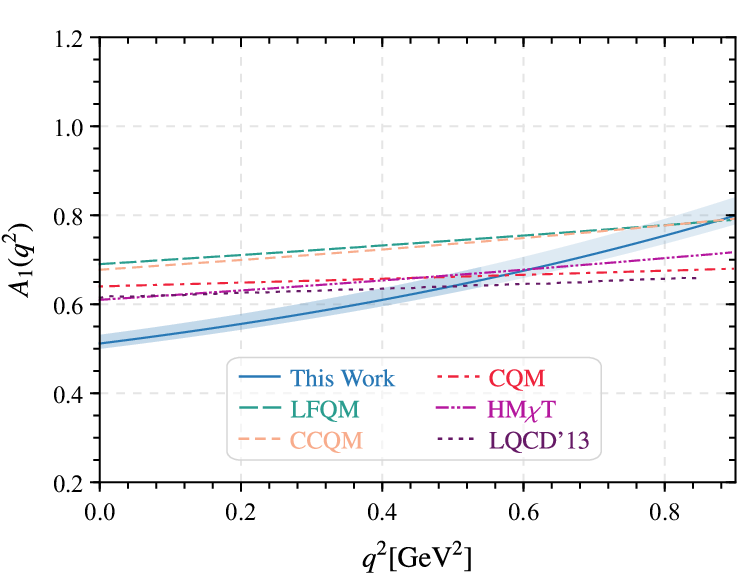}
\includegraphics[width=0.45\textwidth]{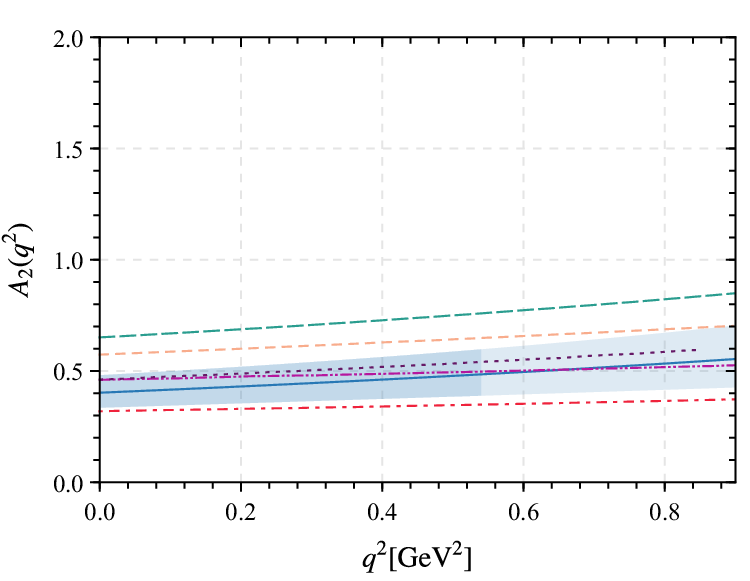}
\includegraphics[width=0.45\textwidth]{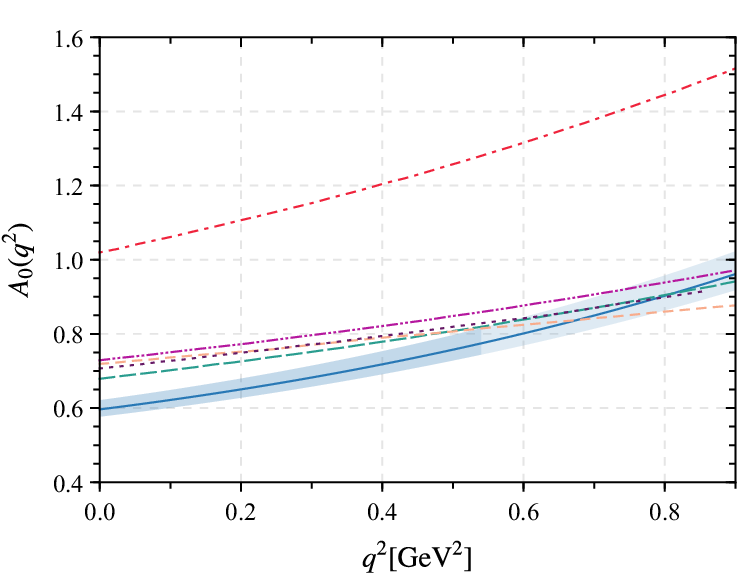}
\includegraphics[width=0.45\textwidth]{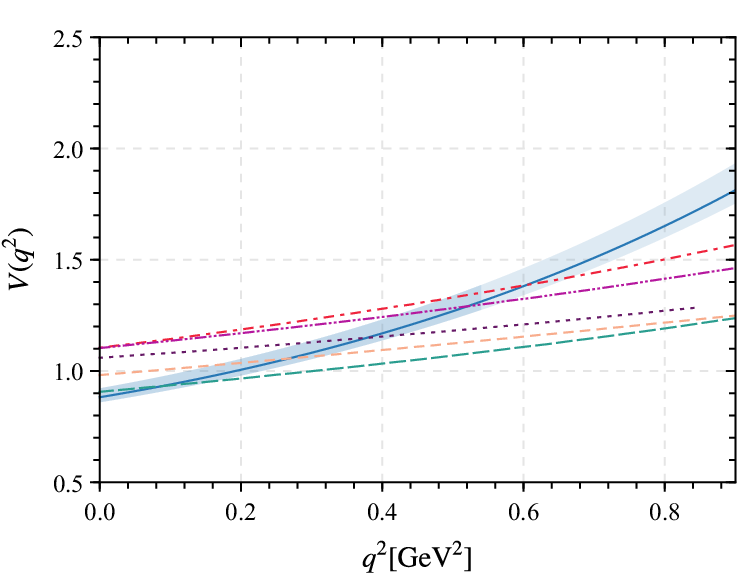}
\end{center}
\caption{The extrapolated TFFs $A_1(q^2)$, $A_2(q^2)$, $A_0(q^2)$ and $V(q^2)$ in the whole $q^2$-region, where the solid line are central values and the shaded bands are corresponding uncertainties. The thicker shaded band shows the LCSR prediction. The ${\rm HM\chi T}$~\cite{Fajfer:2005ug}, the CQM~\cite{Melikhov:2000yu}, the CCQM~\cite{Ivanov:2019nqd}, the CLFQM'11~\cite{Verma:2011yw}, the LQCD'13~\cite{Donald:2013pea} predictions are also presented.}
\label{fig:fq}
\end{figure}

\begin{figure}[htb]
\centering
\includegraphics[width=0.6\textwidth]{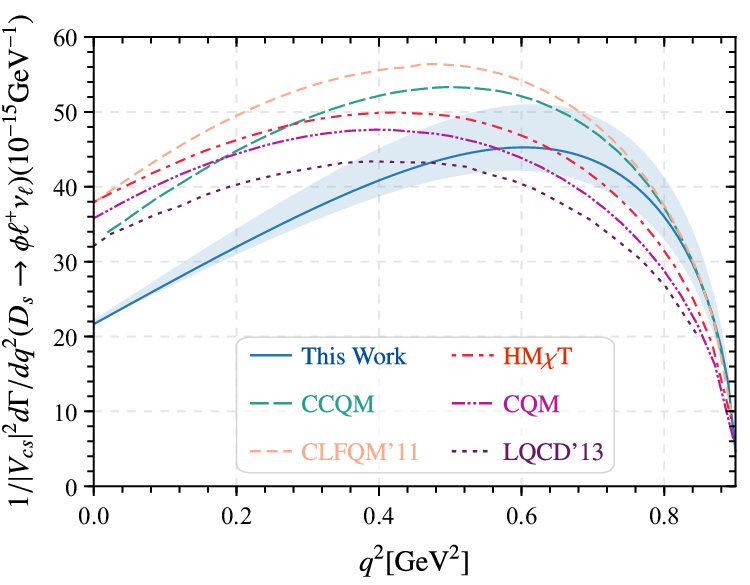}
\caption{The differential decay width $1/|V_{\rm cs}|^2 d\Gamma(D_s^+ \to \phi \ell^+ \nu_\ell)/dq^2$ as a function of $q^2$, where the solid lines are central values and the shaded bands are corresponding uncertainties. The ${\rm HM\chi T}$~\cite{Fajfer:2005ug}, the CQM~\cite{Melikhov:2000yu}, the CCQM~\cite{Ivanov:2019nqd}, the CLFQM'11~\cite{Verma:2011yw}, the LQCD'13~\cite{Donald:2013pea} predictions are also presented.}
\label{Fig:dG}
\end{figure}

Figure~\ref{fig:fq} displays the extrapolated $D_s^+\to\phi$ TFFs across the entire $q^2$ region, where the results under the ${\rm HM\chi T}$~\cite{Fajfer:2005ug}, the CQM~\cite{Melikhov:2000yu}, the CCQM~\cite{Ivanov:2019nqd}, the CLFQM'11~\cite{Verma:2011yw}, the LQCD'13~\cite{Donald:2013pea} predictions are also presented. The shaded bands of our predictions are caused by the input parameters, and the results of other groups are their central predictions. Using those TFFs together with the formula \eqref{Eq:dg}, we calculate the differential decay width $1/|V_{\rm cs}|^2 d\Gamma(D_s^+ \to \phi \ell^+ \nu_\ell)/dq^2$, and the results are presented in Figure~\ref{Fig:dG}. It shows that our differential decay width exhibits a significant deviation from other results in low $q^2$ region. This is reasonable, since the decay width is dominated by the TFF $A_{1}(q^2)$ and we have a smaller $A_1(q^2)$ in $q^2=0$ as shown by Table~\ref{Tab:TFF}. Integrating them over $q^2$ in entire physical $q^2$-region from $m_\ell^2$ to $(m_{D_s^+}-m_\phi)^2$, we then obtain
\begin{eqnarray}
&& \Gamma_{\rm L} = \left(16.914^{+1.382}_{-0.532}\right) \times 10^{-15}~{\rm GeV},  \label{Eq:width1} \\
&& \Gamma_{\rm T} = \left(15.605^{+2.133}_{-1.276}\right) \times 10^{-15}~{\rm GeV},  \label{Eq:width2} \\
&& \Gamma_{\rm total}  = \left(32.519^{+3.515}_{-1.808}\right) \times 10^{-15}~{\rm GeV}. \label{Eq:width3}
\end{eqnarray}
It indicates $\left(\Gamma_{\rm L}/\Gamma_{\rm T} =1.084^{+0.059}_{-0.052}\right)$, which is consistent with the CLEO data $(\Gamma_{\rm L} /\Gamma_{\rm T})_{\rm exp}=1.0\pm0.3\pm0.2$~\cite{CLEO:1994msc} within errors.
\begin{table}[htb]
\renewcommand\arraystretch{1.15}
\centering
\footnotesize
\caption{Typical experimental and theoretical predictions on the $D_s^+ \to \phi\ell^+\nu_\ell$ branching fractions and their corresponding errors (in unit: $10^{-2}$). }
\label{Tab:bf}
\begin{tabular}{lll}
\hline
Experiments~~~~~~~~~~~~~~~~~~~~~& ${\cal B}(D_s^+\to \phi e^+\nu_e) $~~~~~~~~~~~~~~~~~   & ${\cal B}(D_s^+\to \phi \mu^+\nu_{\mu})$
\\
\hline
BESIII'17~\cite{BESIII:2017ikf}     & $2.26\pm0.45\pm0.09$        & $1.94\pm0.54$    \\
BESIII'23~\cite{BESIII:2023opt}     & -                    & $2.25\pm0.09\pm0.07$    \\
CLEO~\cite{Hietala:2015jqa}    & $2.14\pm0.17\pm0.08$         & -       \\
BABAR~\cite{BaBar:2008gpr}   & $2.61\pm0.11\pm0.15$          & -        \\
PDG~\cite{Zyla:2022zbs}     & $2.39\pm0.16$                  & $1.90\pm0.5$                   \\   \hline
Theories& ${\cal B}(D_s^+\to \phi e^+\nu_e) $   & ${\cal B}(D_s^+\to \phi \mu^+\nu_{\mu})$~
\\   \hline
This work   & $2.367_{-0.132}^{+0.256}$   & $2.349_{-0.132}^{+0.255}$  \\
CLFQM'17~\cite{Cheng:2017pcq}    & $3.1\pm0.3$                  & $2.9\pm0.3$                   \\
3PSR'04~\cite{Du:2003ja}     & $1.80\pm0.50$          & -        \\
CLFQM'08~\cite{Wang:2008ci}  & $2.30$          & -        \\
LCSR~\cite{Aliev:2004vf}     & $2.15_{-0.31}^{+0.27}$          & -        \\
HQEFT~\cite{Wu:2006rd}    & $2.53_{-0.40}^{+0.37}$         & $2.40_{-0.40}^{+0.35}$              \\
CCQM~\cite{Ivanov:2019nqd}   & $3.01$                  & $2.85$                    \\
CQM~\cite{Melikhov:2000yu}    & $2.57$                  & $2.57$                   \\
$\chi {\rm UA}$~\cite{Sekihara:2015iha}     & $2.12$        & $1.94$               \\
RQM~\cite{Faustov:2019mqr}   & $2.69$      & -               \\
SCI~\cite{Xing:2022sor}      & $2.45$        & $2.30$               \\
\hline
\end{tabular}
\end{table}

By using the lifetime $\tau_{D_s^+}=0.504~{\rm ps}$ and the CKM matrix element $V_{cs}=0.975$~\cite{Zyla:2022zbs}, we obtain the branching fractions for $D_s^+\to\phi\ell^+\nu_{\ell}$ with $\ell=(e,\mu)$, which are presented in Table~\ref{Tab:bf}. For ${\cal B}(D_s^+\to \phi e^+\nu_e)$, our result agrees well with the PDG averaged value~\cite{Zyla:2022zbs} and are consistent with other data within errors. For ${\cal B}(D_s^+\to \phi e^+\nu_e)$, our prediction fall within the error range reported by the recent more precise BESIII'23 data~\cite{BESIII:2023opt}. By using the world average of ${\cal B}(D_s^+\to \phi e^+\nu_e)$, the BESIII group then issued the ratio of those two branching fractions, {\it e.g.} $\frac{{\cal B}(D_s^+\to \phi \mu^+\nu_\mu)}{{\cal B}(D_s^+\to \phi e^+\nu_e)}=0.94\pm0.08$~\cite{BESIII:2023opt}. Our predicted value $\simeq 0.99$ falls within this margin of error, aligning with the lepton universality.

\subsection{Polarization and asymmetry parameters of $D_s^+\to \phi\ell^+\nu_\ell$}
\begin{figure}[htb]
\centering
\includegraphics[width=0.45\textwidth]{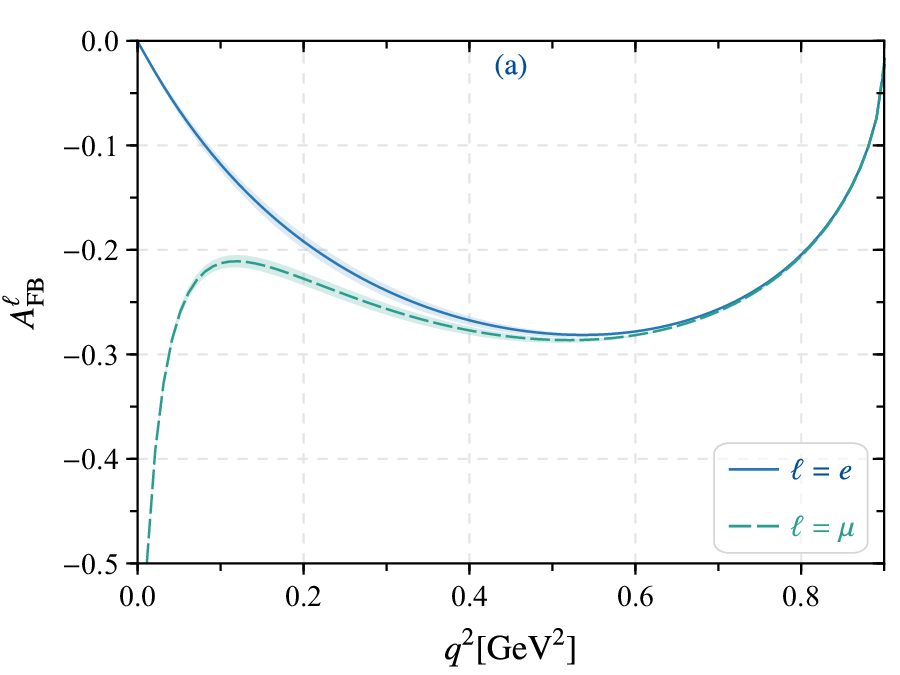}
\includegraphics[width=0.45\textwidth]{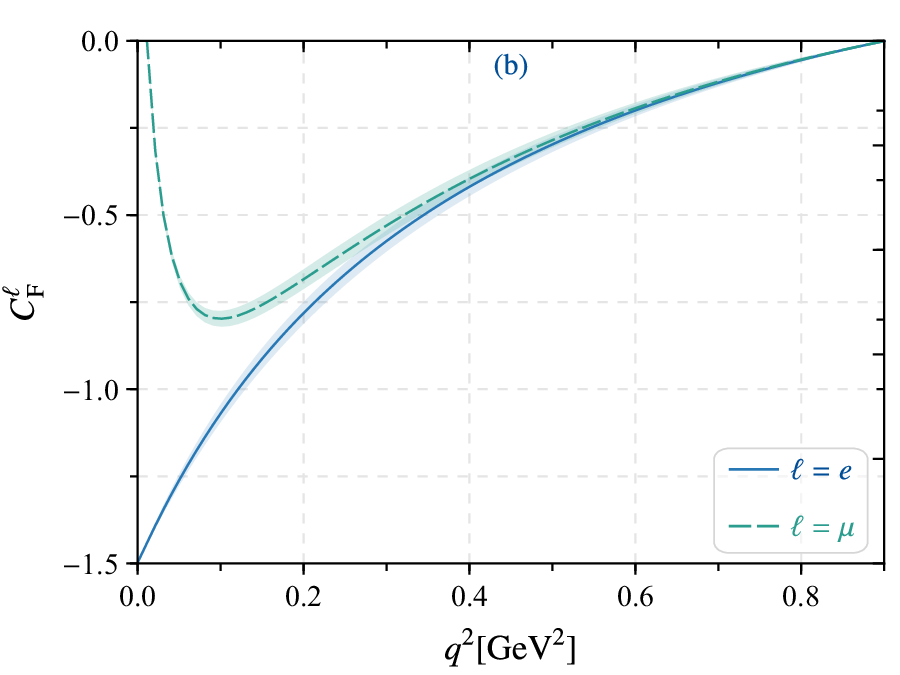}
\includegraphics[width=0.45\textwidth]{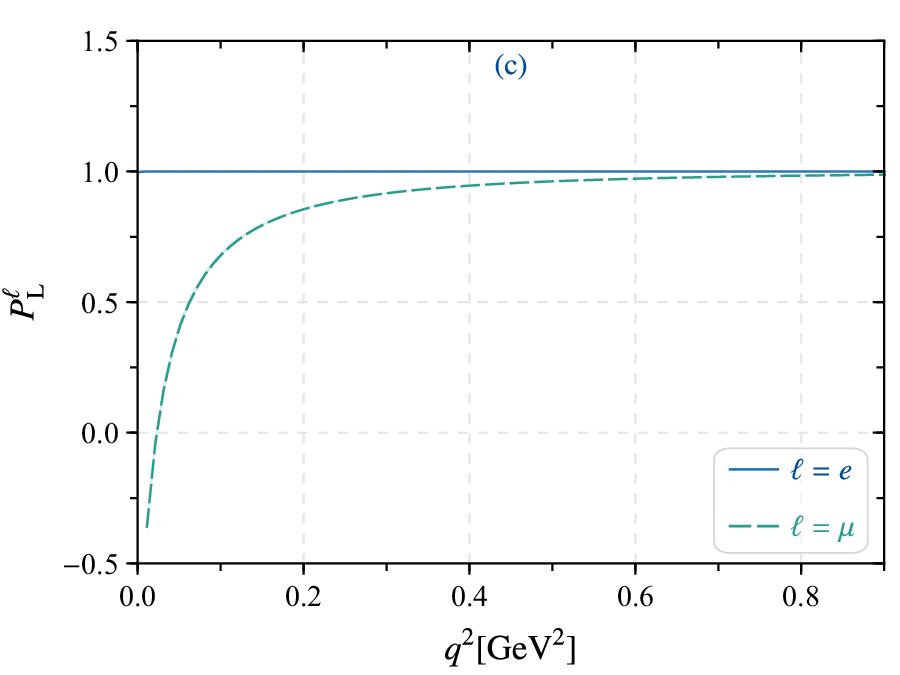}
\includegraphics[width=0.45\textwidth]{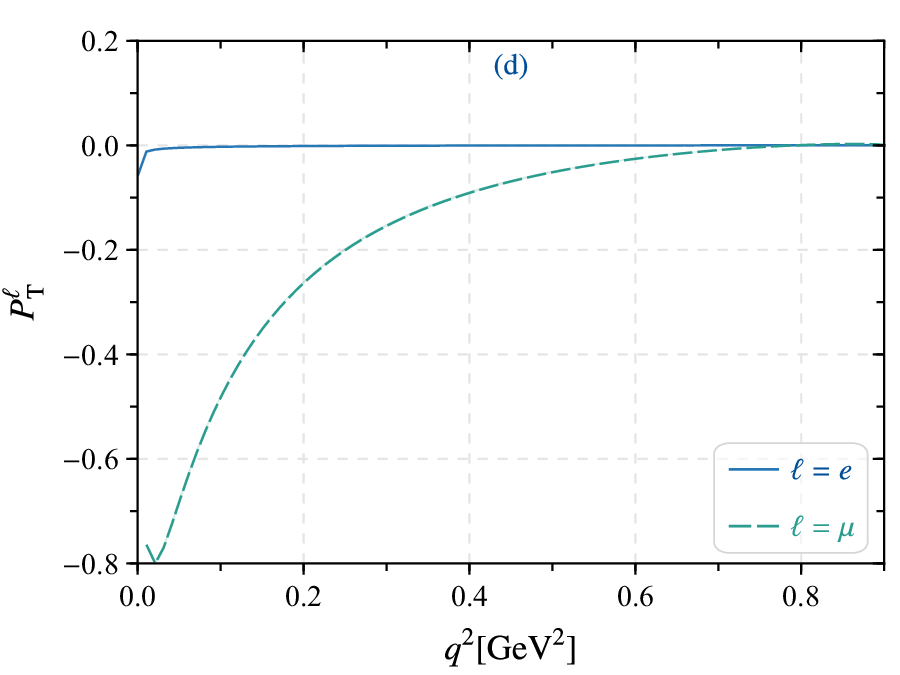}
\includegraphics[width=0.45\textwidth]{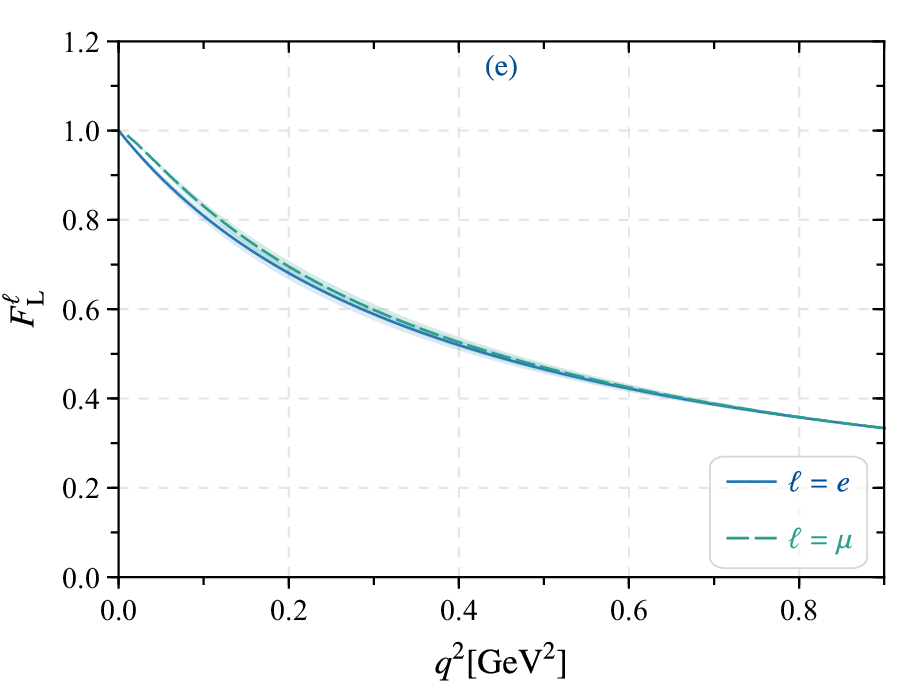}
\caption{The polarization and asymmetry parameters for the semi-leptonic decay $D_s^+\to\phi\ell^+\nu_{\ell}$, including the central value and its uncertainty. The solid and dashed lines show the central values and the shaded bands show the corresponding uncertainties.}
\label{Fig:Newphy}
\end{figure}

In this subsection, we give our prediction for the polarization and asymmetry parameters of $D_s^+\to \phi\ell^+\nu_\ell$. Those observables provide some more detailed information for the semi-leptonic decays of hadron. Substituting the extrapolated TFFs $V(q^2)$ and $A_{0,1,2}(q^2)$ into Eqs.~(\ref{sys1}) - (\ref{sys5}), we can obtain the polarization and asymmetry parameters for $D_s^+\to\phi\ell^+\nu_{\ell}$. We present the polarization and asymmetry parameters $A^{\ell}_{FB}$, $C^{\ell}_{F}$, $P^{\ell}_{L}$, $P^{\ell}_{T}$ and $F^{\ell}_{L}$ versus the $q^2$ in Figure~\ref{Fig:Newphy}, whose errors caused by different choices of input parameters are shown by shaded bands. Figure~\ref{Fig:Newphy} shows that all the uncertainties are small, especially for the case of longitudinal (transverse) polarization of the final charged lepton $P^{\ell}_{L(T)}(q^2)$, whose uncertainties are about thousandth and are negligible.
\begin{table}[htb]
\renewcommand\arraystretch{1.15}
\centering
\footnotesize
\caption{The polarization and asymmetry parameters for the semi-leptonic decay $D_s^+\to\phi\ell^+\nu_{\ell}$. All the input parameters are set to be their central values. } \label{Tab:Newphy}
\begin{tabular}{lllllllllll}
\hline
~~~~  & ~$A^e_{FB}$~  & ~$A^{\mu}_{FB}$~  & ~$C^e_{F}$~   & ~$C^{\mu}_{F}$~  & ~$P^e_{L}$~   & ~$P^{\mu}_{ L}$~  & ~$P^e_{T}$~   & ~$P^{\mu}_{T}$~ & ~$F^e_{L}$~   & ~$F^{\mu}_{L}$~  \\   \hline
This work  & $-0.190$   & $-0.220$   & $-0.421$   & $-0.320$  & $0.900$   & $0.780$    & $-0.001$
& $-0.139$   & $0.487$   & $0.483$      \\
RQM~\cite{Faustov:2019mqr}  & $-0.21$      & $-0.24$   & $-0.49$   & $-0.35$    & $1.00$   & $0.90$    & $0.00$     & $-0.15$   & $0.54$   & $0.54$     \\
CCQM~\cite{Ivanov:2019nqd}  & $-0.18$      & $-0.21$   & $-0.43$   & $-0.34$   & $1.00$   & $0.91$    & $-0.11$    & $-0.14$   & $0.53$   & $0.50$      \\
\hline
\end{tabular}
\end{table}

Integrating the formulas (\ref{sys1}, \ref{sys2}, \ref{sys3}, \ref{sys4}, \ref{sys5}) over $q^2$, we get the integrated values for those observables. We present them in Table~\ref{Tab:Newphy}, where the results derived from the RQM~\cite{Faustov:2019mqr} and the CCQM~\cite{Ivanov:2019nqd} approaches have also been presented. The polarization and asymmetry parameters exhibit variations across different lepton masses. Our predicted $A^{\ell}_{FB}$, $C^{\ell}_{F}$ and $P^{\mu}_T$ fall within the range of RQM and CCQM predictions, while both $P^{\ell}_{L}$ and $F^{\ell}_{L}$ exhibit smaller values compared to the predictions from RQM and CCQM.

\section{Summary}\label{Sec_IV}

In this paper, we have calculated the TFFs for the semi-leptonic decay $D_s^+\to\phi\ell^+\nu_{\ell}$ by using the LCSR approach. Numerical results for those TFFs and two typical ratios $\gamma_V$ and $\gamma_2$ at the large recoil point $q^2=0$ have been given in Table~\ref{Tab:TFF} and Table~\ref{Tab:ratio}. In doing the calculation, we have suggested an improved the LCHO model for the leading-twist LCDA of $\phi$-meson, whose $\xi$-moments $\langle\xi^{\|;n}_{2;\phi}\rangle$ can be determined by using the QCD SR approach within the background field theory. And to improve its accuracy, these moments have been calculated up to $10_{\rm th}$-order accuracy. A comparison of various LCDA models at the $2$ GeV has been shown in Figure~\ref{fig:phi}. Our model exhibits a single-peak behavior, closely resembling the conventional asymptotic form for the light-mesons and the one suggested from the Lattice QCD calculation.

W have presented the differential decay width of the semi-leptonic decay $D_s^+\to\phi\ell^+\nu_{\ell}$ with $\ell=(e,\mu)$ in Figure~\ref{Fig:dG} and the branching fractions in Table~\ref{Tab:bf}. For both the electron and muon channels, our predicted branching fractions are consistent with the experimental data within errors. Additionally, we have calculated its longitudinal, transverse and total decay widths, which are shown by Eqs.(\ref{Eq:width1}, \ref{Eq:width2}, \ref{Eq:width3}). They indicates that $\left(\Gamma_{\rm L}/\Gamma_{\rm T} =1.084^{+0.059}_{-0.052}\right)$, which is consistent with the CLEO data~\cite{CLEO:1994msc} within errors. Using those results, we have also estimated the forward-backward asymmetries, the lepton-side convexity parameters, as well as the lepton and vector meson longitudinal and transverse polarization parameters, which have been collected in Table~\ref{Tab:Newphy} and Figure~\ref{Fig:Newphy}. Those values can be measured and tested in future experiments, which could be inversely adopted for testing the various $\phi$-meson LCDA models.

\section{Acknowledgments}

This work was supported in part by the National Natural Science Foundation of China under Grant No.12347101, 12265010, 12265009 and No.12175025, the Project of Guizhou Provincial Department of Science and Technology under Grant No.ZK[2021]024, the Graduate Research and Innovation Foundation of Chongqing, China under Grant No.CYB23011 and No.ydstd1912, the Fundamental Research Funds for the Central Universities under Grant No.2020CQJQY-Z003.

\end{document}